\documentclass[review,nonatbib]{elsarticle}

\usepackage{amsmath}
\usepackage{amssymb}
\usepackage{crimson}
\usepackage[
a4paper, margin=1in
]{geometry}
\usepackage{graphicx} 
\usepackage{url}
\usepackage{subcaption}
\usepackage{caption}
\usepackage{floatrow}
\usepackage{booktabs}
\usepackage{hyperref}
\hypersetup{colorlinks=true}

\makeatletter
\let\c@author\relax
\makeatother

\usepackage[url=false,sorting=none,citestyle=numeric-comp]{biblatex}
\AtEveryBibitem{\clearfield{note} \clearlist{language}}
\usepackage{threeparttable}
\usepackage{setspace}
\journal{ArXiv}

\begin{document}
\begin{abstract}

Long-duration energy storage (LDES) is a key component for fully renewable, sector-coupled energy systems based on wind and solar. While capacity expansion planning has begun to take into account interannual weather \textit{variability}, it often ignores weather \textit{uncertainty} and limited foresight in capacity and operational decisions. We build a stochastic capacity expansion model for fully decarbonized energy systems with LDES in Europe accounting for weather uncertainty - isolating the effect of limited foresight by comparing it to a perfect foresight benchmark. Under limited foresight, LDES acts as a hedge against extreme system states operating defensively and exhibiting a stockpiling effect absent under perfect foresight. Solar PV gains in system value for its higher predictability with up to 25\% higher capacities versus the benchmark while onshore wind capacities are lower. We shed light on the underlying mechanisms by deriving implicit LDES bidding curves. We show that LDES bids reflect the costs and the weather-dependent probability of extreme system states conditional on the current system state. This has important implications for the price formation on renewable electricity markets, as a wide and continuous range of probabilistic LDES bids alleviates concerns of extreme price disparity at high renewable shares. 
\end{abstract}


\begin{frontmatter}

\title{On long-duration storage, weather uncertainty and limited foresight}

\author[A1,A2]{Felix Schmidt\corref{FS}}
\ead{fschmidt@diw.de}

\address[A1]{DIW Berlin, Department of Energy, Transportation, Environment, Mohrenstra{\ss}e 58, 10117 Berlin, Germany}
\address[A2]{Technische Universität Berlin, Straße des 17. Juni 135, 10623 Berlin, Germany}
\end{frontmatter}

\section{Introduction}
The energy sector is at the center of climate change mitigation efforts around the world \cite{international_energy_agency_net_2024,intergovernmental_panel_on_climate_change_ipcc_technical_2023}. Variable renewable energy (VRE), including wind and solar energy, is likely to supply the majority of the energy in a decarbonized system for its ubiquitous availability, cost competitiveness and ease of deployment \cite{schlachtberger_cost_2018,victoria_speed_2022,brown_response_2018,breyer_history_2022,luderer_impact_2022,hansen_status_2019}. The resulting system dependency on weather variables such as wind speeds, solar irradiation, water inflows and temperatures due to electrified heating and cooling demand creates flexibility requirements on various temporal and spatial scales \cite{lund_review_2015,levin_energy_2023}.

Long-duration energy storage (LDES) is considered a key enabler of very high VRE shares in a decarbonized future by providing seasonal or interannual balancing and bridging periods of low solar and wind feed-in that may last multiple days or even weeks, so-called \textit{Dunkelflauten} \cite{lopez_prol_economics_2021,dowling_role_2020,ruhnau_storage_2022,sepulveda_design_2021,kittel_coping_2025,staadecker_value_2024}. LDES is not well defined but often used to refer to storage technologies that offer durations between ten and hundreds of hours and which are characterized by low energy capacity costs \cite{dowling_long-duration_2021,albertus_long-duration_2020,brown_ultra-long-duration_2023}. While several technology options exist, most of the academic literature, planned legislation and current market developments in Europe and the US suggest that hydrogen storage in salt caverns in conjunction with electrolysis and hydrogen gas turbines (power-to-gas-to-power) will play a dominant role \cite{dowling_role_2020,federal_ministry_for_economic_affairs_and_climate_action_of_germany_weisbuch_2025,rwe_erweiterungsprojekt_2024,uniper_press_2024}.

Capacity expansion models (CEM) are well-established and important tools for policymakers, investors and other stakeholders in the energy transition. Yet, the representation of LDES in these models is inherently difficult \cite{levin_energy_2023}, mainly for two reasons: 

\textit{First}, while the literature has reached a broad consensus that VRE-based energy system models need to capture the spatio-temporal distribution of weather variables to find cost-effective and resilient system designs \cite{collins_impacts_2018,schlachtberger_cost_2018,staffell_increasing_2018,gotske_designing_2024,goke_stabilized_2024,grochowicz_intersecting_2023}, LDES is \textit{particularly} sensitive to the interannual variability in these variables \cite{dowling_long-duration_2021,ruhnau_storage_2022,kittel_coping_2025}, which is linked to the fact that LDES derives its system value from extreme system states happening rather infrequently. Recent literature contributions have sought to make the inclusion of a large number of weather years at a high granularity computationally feasible \cite{grochowicz_intersecting_2023,goke_stabilized_2024,jacobson_computationally_2024}. However, most studies determine optimal LDES capacities based on a single weather year, potentially leading to highly sub-optimal capacity choices when evaluated against other samples from the (unknown) weather distribution \cite{zeyringer_designing_2018}. 

\textit{Second}, CEMs, even those taking into account interannual variability by including multiple weather years, commonly assume \textit{perfect foresight} in the dispatch stage. This is another practical concession made to maintain computational tractability in light of high spatial, temporal and technological granularity commonly found in these models \cite{conejo_generation_2016}. Assuming perfect foresight is particularly consequential for the operation of LDES. For example, LDES injection decisions in summer would be taken under perfect knowledge of the looming VRE scarcity events and necessary storage withdrawal in winter, which may lead to very unrealistic LDES dispatch patterns, LDES system value and consequently capacity decisions. In reality, a LDES or system operator would only have \textit{limited foresight} based on (probabilistic) weather forecasts. 

Some studies have introduced heuristic decision rules to eliminate effectively foresight in LDES operations \cite{dupre_la_tour_towards_2023,brown_price_2025}. These rules impose to sell and buy energy at fixed prices or determine bids as a function of the current storage level. These approaches are likely to underestimate LDES system value and may bias capacity decisions. By contrast, the literature on hydroelectric power systems has long dealt with optimal storage dispatch under inflow uncertainty \cite{scott_modelling_1996,steeger_optimal_2014}. A fully renewable energy system based on wind and solar and relying on LDES for balancing is in many respects similar to a hydro-dominated system \cite{blanchard_strategic_2024}, where random water inflows are replaced by random feed-in of wind and solar irradiation. 

We build on this insight and adopt a prominent solution algorithm for the resulting multi-stage stochastic programs from the hydro literature, Stochastic Dual Dynamic Programming (SDDP) \cite{pereira_multi-stage_1991}, which has recently been extended to include capacity decisions in an energy system with VRE generation \cite{hole_capacity_2025,blanchard_strategic_2024}.

This paper has three main contributions: \textit{first}, we present a capacity expansion model for fully renewable, wind- and solar-based European energy systems subject to both demand-side (electrified heating) and supply-side (variable renewable energy) weather uncertainty. \textit{Second}, we isolate the effect of limited foresight by comparing the multi-stage formulation to a two-stage formulation that takes into account interannual weather variability but assumes perfect foresight in the dispatch stage. We investigate differences in LDES operations and consequences for optimal capacity decisions. \textit{Third}, we derive LDES bidding curves implicit in the optimal policy of the multi-stage formulation and discuss their dependence on the cost of backup or loss of load. We also show that LDES bids under weather uncertainty have important implications for the price formation on fully renewable energy-only markets and thereby contribute to the current debate on the viability of energy-only markets without fuel costs \cite{brown_price_2025,antweiler_new_2025}.


\section{Capacity expansion planning of renewable energy systems under weather uncertainty} \label{sec:theory}

The sensitivity of VRE-based energy systems to interannual weather variability and extreme weather events is well documented \cite{pfenninger_dealing_2017,zeyringer_designing_2018,staffell_increasing_2018,bloomfield_quantifying_2021,craig_overcoming_2022,grochowicz_using_2024}. Consequently, it has become common practice to include more than one 'typical meteorological year' in capacity expansion models \cite{grochowicz_using_2024}. However, taking into account interannual variability does not necessarily account for weather \textit{uncertainty} even if the terms are often conflated in this context. Weather \textit{uncertainty} means that the realization of relevant weather variables in the future is not known to the decision maker by the time they make their decision and the problem becomes \textit{stochastic} \cite{conejo_decision_2010}. In other words, the decision maker has \textit{limited foresight} with respect to \textit{random} weather variables. A deterministic CEM, not accounting for uncertainty, has a single decision stage and the decision maker decides on capacity and dispatch at once. In a \textit{stochastic} CEM, information arrives and decisions are made in sequence.

Broadly, there are four ways in which the literature has included multiple weather years in their analyses. The first two account for weather variability but are deterministic. The latter two are stochastic CEM formulations accounting for weather uncertainty.

\begin{figure}[!htp]
    \centering
    \includegraphics[width=0.8\linewidth]{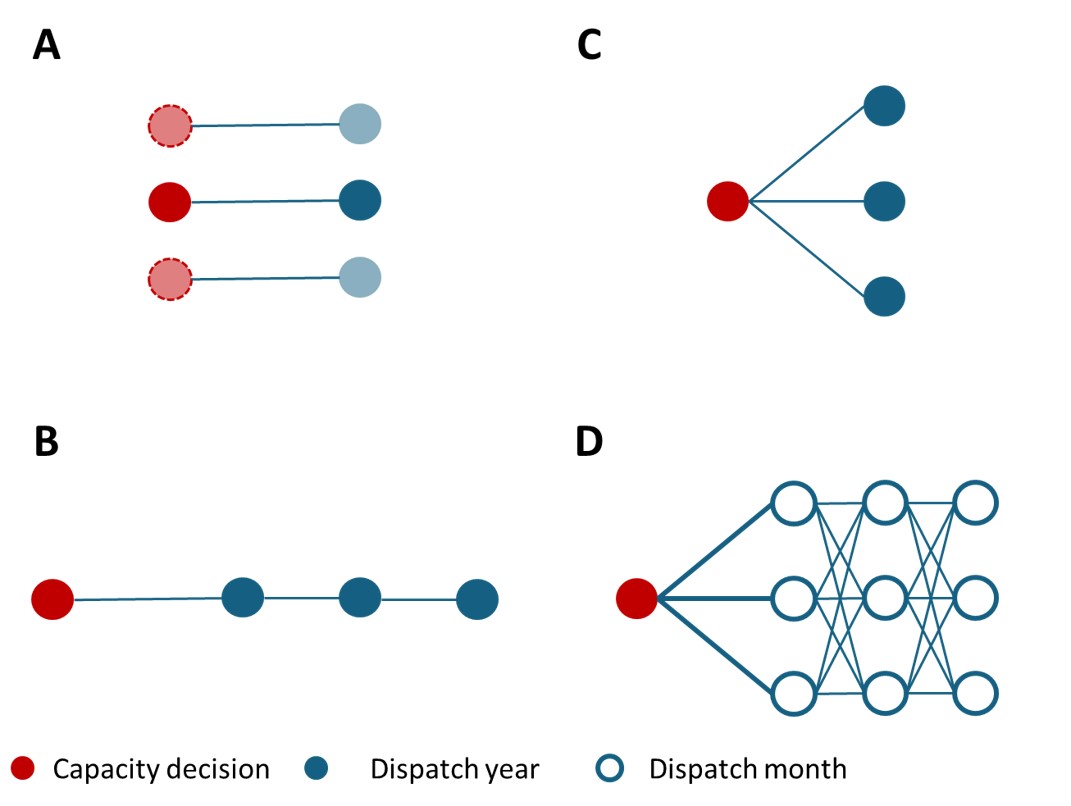}
    \caption{\textit{Model setups to deal with weather variability and uncertainty}: Panel A represents single-year deterministic models. Panel B is a deterministic model that includes a (long) sequence of weather years in the dispatch stage. Panel C is a two-stage stochastic setup in which the capacity decision is taken under uncertainty. There is perfect foresight in the dispatch stage. Lastly, Panel D represents the multi-stage stochastic setup that limits foresight in the dispatch stage.}
    \label{fig:types}
\end{figure}

Panel A of Figure \ref{fig:types} represents the most common way, in which the joint optimization of capacity and dispatch is repeated for every weather year. Resulting in a range of model outputs, this type can be considered a \textit{sensitivity analysis}  \cite{levin_energy_2023}. In the stochastic programming literature, this approach corresponds to a \textit{wait-and-see} solution \cite{shapiro_lectures_2009}. Some analyses have added a second step in which the system design based on one weather year is tested for dispatch on another year to find resilient system specifications among all output candidates \cite{gotske_designing_2024,ruggles_planning_2024}.

Panel B of the same figure represents contributions that solve the dispatch stage of a CEM over potentially long sequential periods of up to 70 weather years \cite{brown_ultra-long-duration_2023}. Especially used in the context of LDES to capture interannual storage dynamics \cite{zeyringer_designing_2018,dowling_role_2020,ruhnau_storage_2022}, these models make capacity decisions given many weather years but are potentially distorted by extreme levels of perfect foresight. For example, Brown and Hampp \autocite{brown_ultra-long-duration_2023} show storage level trajectories that imply an overarching discharge period of nearly a decade.

Panel C represents cases where each dispatch year is optimized in isolation but the capacity mix is constrained to be the same across weather years. In stochastic programming, this corresponds to a \textit{two-stage} stochastic program \cite{shapiro_lectures_2009}. Recent literature contributions have sought to make models of this type computationally tractable \cite{goke_stabilized_2024,grochowicz_intersecting_2023,jacobson_computationally_2024}. While they only differ from Panel B in that they do not consider dispatch in sequence, it represents a capacity decision under weather uncertainty, albeit with perfect foresight during dispatch.

Panel D shows a multi-stage stochastic capacity expansion model \cite{shapiro_lectures_2009} abandoning the perfect foresight assumption during the dispatch stage. Existing literature contributions are mainly found for hydro-dominated systems, e.g. \cite{rebennack_generation_2014,thome_stochastic_2019,hole_capacity_2025}, some of which consider stochastic wind and solar generation next to random inflows. While Blanchard \cite{blanchard_strategic_2024} and Hummelen et al. \cite{hummelen_exploring_2024} explicitly consider hydrogen storage with limited foresight under solar and wind uncertainty, they only consider a partial capacity choice between electrolyzers and storage to hedge against hydrogen supply disruptions or use a very reduced scenario tree, respectively.

Other treatments of (weather) uncertainty include robust or chance-constrained optimization approaches. Roald et al. \cite{roald_power_2023} provide a comprehensive treatment of different representations of uncertainty in power system models.

This work isolates the impact of perfect foresight on LDES operations and the for optimal capacity decisions in fully renewable wind- and solar-based energy systems by comparing the outcomes of model types D and C. The remainder of this section introduces formally a CEM of type D and discusses necessary assumptions and features of the resulting model. We use the terms system planner and system operator interchangeably.



\subsection{Modeling capacity expansion with limited weather foresight}

A CEM of type D leads to a multi-stage stochastic program (MSSP), where the first stage $t=0$ corresponds to the capacity decision and stages $t=1,\dots,T$ are dispatch stages. Each dispatch stage consists of $H(t) = |\mathcal{H
}(t)|$ operational periods. When in dispatch stage $t$, the system operator has perfect foresight for all operational periods $h\in\mathcal{H}(t)$ in the stage, $\mathcal{H
}(t)$. Weather realizations beyond $t$ are uncertain and the system operator needs to take the distribution of future outcomes into account when making a decision in stage $t$. Generally, this leads to a sequence of nested expectations, where each expectation is conditional on the past decisions. We provide a general problem formulation in Section \ref{sec:si_model_description}.

The solution to such an MSSP is a \textit{policy}, a set of functions that maps system states to capacity or dispatch choices. A system state is described by two components: 1) the exogenous weather realization, and 2) an endogenous state resulting from previous decisions, such as capacity choices from stage $0$ or the LDES storage level inherited from the previous dispatch stage. 

The Bellman Principle of Optimality \cite{bellman_theory_1952} ensures that a policy solving the multi-stage CEM satisfies in every dispatch stage $t\in \{1,\dots, T\}$,
\begin{subequations} \label{eq:Bellman}
\begin{equation}
 \begin{aligned}
     Q_t(e_{s,t-1,H},G_r,F_s,H_s,E_s,e^{ini}_s,\xi_{t}):=& \\\min_{\substack{\{g_{r,t,h},f_{s,t,h},h_{s,t,h},g^{inf}_{t,h},e_{s,t,h}\} \\ \in \mathcal{O}_t(G_r,F_s,H_s,E_s,e_s^{ini},e_{s,t-1,H},\xi_{t})}}  \sum_h\left[\sum_r o_r g_{r,t,h} +vg^{inf}_{t,h}\right] &+ \mathbb{E}_t\left[{Q}_{t+1}(e_{s,t,H},G_r,F_s,H_s,E_s,e_s^{ini},\xi_{t+1})\right]
\end{aligned}
\end{equation}

The optimal value function $Q_t(\cdot)$ is a function of state variables, where $G_r,F_s,H_s,E_s$ are decisions for generation capacities, storage power and storage energy capacities from stage $0$ respectively. $e_{s,t-1,H}$ is the storage ending level of storage $s$ in the previous stage $t-1$, $e_s^{ini}$ is the initial storage level that needs to be met again at the end of stage $T$ (see Section \ref{sec:storage_target}) and $\xi_t\in\Xi_t$ is the random weather vector, which contains random capacity factors for wind and solar PV, hydro flows, heating demand and heat pump efficiencies for every period $h\in \mathcal{H}(t)$. These state variables affect the feasible space for the dispatch decisions in $t$, $\mathcal{O}_t(\cdot)$. The subproblem in $t$ is to minimize the sum of the stage's dispatch costs, given by the sum of variable generation costs $\sum_ro_rg_{r,t,h}$ and the costs of load shedding $vg_{t,h}^{inf}$ in every period and the \textit{expected} future system costs, the \textit{cost-to-go}. The latter are affected by the decisions in $t$ through the ending storage level $e_{s,t,H}$. The trade-off boils down to reducing the current stage's dispatch costs at the expense of increasing the cost-to-go as the future is left with a lower storage level $e_{s,t,H}$. In stage $T$ the cost-to-go term is equal to $0$. We introduce a terminal condition to avoid greedy storage discharging, which is discussed in Section \ref{sec:storage_target}. 

In the capacity stage $t=0$ the subproblem reads,
\begin{equation}
\begin{aligned}
      &\min_{\{G_r,F_s,H_s,E_s,e_s^{ini}\}\in\mathcal{C}} \sum_r c_rG_r + \sum_s\left( c_s^f F_s +c_s^hH_s +c_s^eE_s\right) + \mathbb{E}_0\left[ Q_1(G_r,F_s,H_s,E_s,e_s^{ini},\xi_1) \right]
\end{aligned}
\end{equation}

\end{subequations}

\noindent where $\mathcal{C}$ is the feasible set for capacity decisions and $c_r,c_s^f,c_s^h$ and $c_e^s$ are unit annualized capacity investment costs.

The expectation in stage $t$ of Problem \ref{eq:Bellman} is with respect to the distribution of $\xi_{t+1}$, which we assume to be independent of $\xi_t$ and any stages prior to $t$. Section \ref{sec:stagewise} discusses this \textit{stagewise independence} assumption in detail. Even with this simplifying assumption, the nested structure of the described problem makes it very hard, if not intractable, to solve for the well-known \textit{curse of dimensionality} \cite{shapiro_analysis_2011}.

We assume that the sample space for each stage $\Xi_t$ is finite and the number of possible weather realizations in stage $t$ is given by $N_t$. Even when $N_t$ are small, the number of possible trajectories of the stochastic process $\xi=\{\xi_t\}_{t=1}^T \in \Xi=\Xi_1\times\cdots\times\Xi_T$ explodes quickly \cite{shapiro_analysis_2011}.\footnote{Assume monthly stages and four possible realizations per month. The size of the sample space would be $4^{12} \approx 16.7 \text{ million}$.} A solution of the monolithic problem is intractable.  

Stochastic Dual Dynamic Programming (SDDP) \cite{pereira_multi-stage_1991} is an algorithm that can circumvent the curse of dimensionality and solve the MSSP approximately by combining Monte Carlo sampling and Benders decomposition methods. 

Broadly, SDDP decomposes the nested problem by replacing the cost-to-go functions $\mathcal{Q}_{t+1}(\cdot):=\mathbb{E}_t[Q_{t+1}(\cdot)] = N^{-1}_t \sum_{i} Q_{t+1}(\cdot,\xi_{t+1,i})$ by cutting plane approximations $\hat{\mathcal{Q}}_{t+1}$, which are defined by a collection of linear constraints, so-called Benders cuts. Each subproblem is now a small linear program (LP) that is easy to solve on its own.

SDDP improves the approximations $\{\hat{\mathcal{Q}}_{t}\}$ iteratively using a forward and a backward pass. A forward pass begins with sampling a weather year path from $\Xi$. Next, each subproblem is solved in sequence given the sampled weather year and the current cost-to-go function approximations. The resulting endogenous state variables, capacities and storage levels, are passed from one subproblem to the next. 

In a subsequent \textit{backward pass} the algorithm moves through the stages in reverse order. Conditional on the incoming endogenous state variables defined in the forward pass, it collects feedback from each subproblem in the form of Benders cuts. These cuts improve the cost-to-go function approximation used in the preceding stage, respectively. We defer a detailed description of the algorithm to the Supplemental Information, Section \ref{sec:sddp}. 

Ultimately, a set of cost-to-go functions $\{\hat{\mathcal{Q}}^K_t\}_{t=1}^T$ implies an approximately optimal policy after $K$ iterations, with which we can simulate (approximately) optimal behavior under limited foresight. $K$ may be predetermined or the result of some other stopping rule. We discuss termination criteria in Section \ref{sec:implementation} 

The perfect foresight benchmark model is of type C. We provide additional details in the Supplemental Information, Section \ref{sec:si_model_description}. We refer to the two models as Limited Foresight (LF) and Perfect Foresight (PF), respectively.

\subsection{Foresight and stagewise independence} \label{sec:stagewise}

The multi-stage stochastic program formulation of a capacity expansion model presented in the last section did not specify the number of stages $T$, or the stage lengths $H(t)$. Any investigation of the effect of limited foresight on renewable energy systems naturally begs the question of how \textit{limited} foresight actually is. The model implies that the system planner has perfect foresight for the duration of each stage.

\begin{figure}[H]
    \centering
    \includegraphics[width=\linewidth]{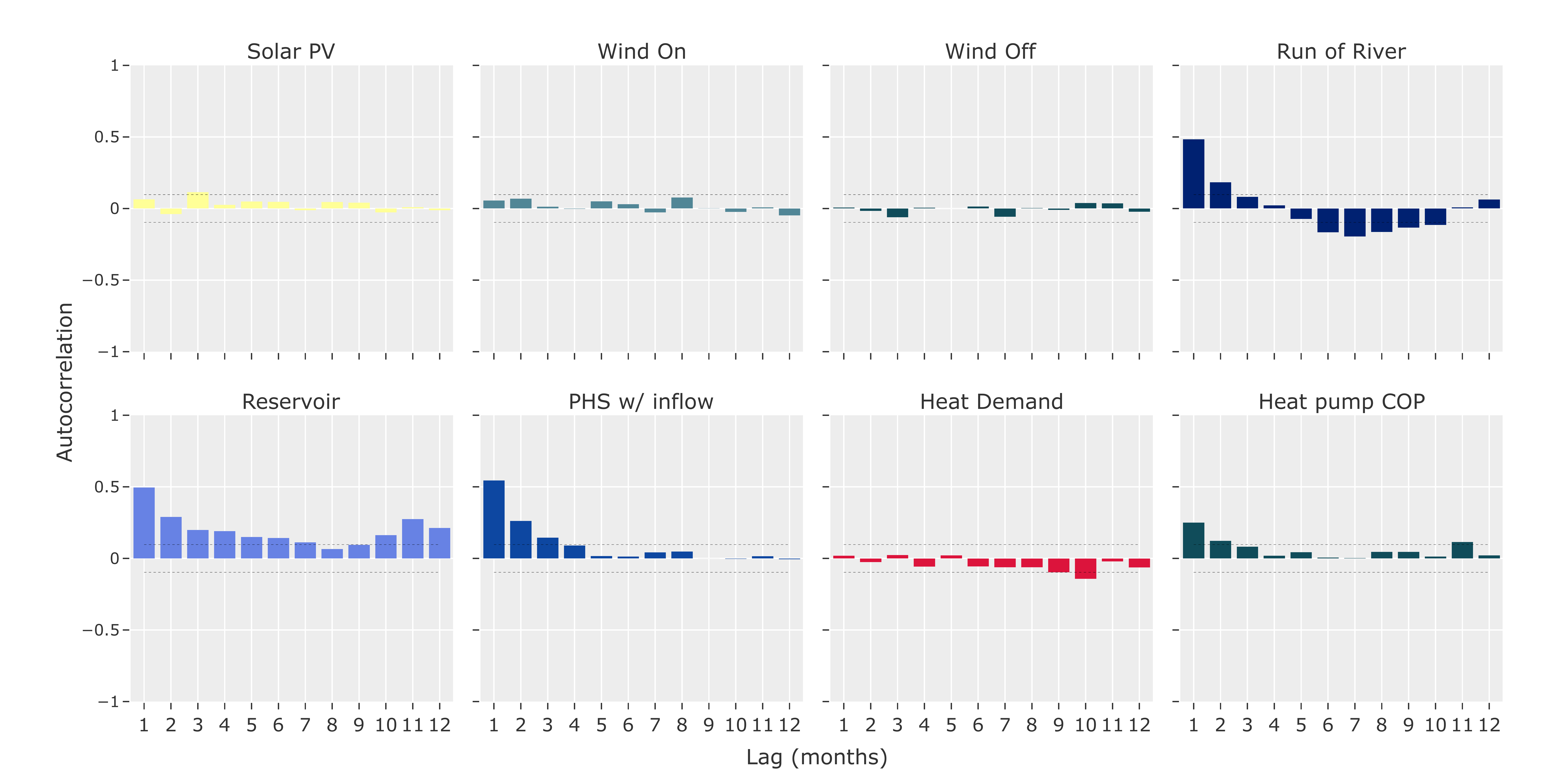}
    \caption{\textit{Autocorrelation functions of weather variables at monthly stage length}. Monthly autocorrelation for solar photovoltaic, wind onshore, wind offshore, run of river, hydro reservoirs, pumped hydro storage (PHS) with inflows, residential and commercial aggregate heat demand and heat pump efficiencies.}
    \label{fig:acf_monthly}
\end{figure}

In the face of weather uncertainty, this is directly related to the system planner's ability to produce or procure reliable weather forecasts. Currently, mid-latitude weather forecasts are reliable up to ten days out. While meteorologists see room for improvements, some have found an intrinsic, that is, theoretical limit to be 14 days in advance \cite{bauer_quiet_2015,selz_transition_2022}. Broader developments could eventually be predicted a month ahead, as a joint research program by the University of Reading, the UK Met Office and the European Center for Medium-Range Weather Forecasts suggests \cite{national_centre_for_atmospheric_science_how_2024}. 

Brown et al. \cite{brown_price_2025} opt for a 96-hour forecast horizon. The literature on hydroelectric energy systems typically assumes stage lengths from a week up to a month \cite{hole_capacity_2025,shapiro_risk_2013}. Yet, the forecast for hydrological inflows may be different from that for solar irradiation and wind speeds. 

We choose a rather long stage length of one month for two reasons. For one, as we explain below, a monthly stage length is consistent with the stagewise independence assumption we made in the previous section. For another, assuming a longer foresight horizon is likely to be conservative in that it underestimates the effect of weather uncertainty on capacity and dispatch decisions in a CEM. Shorter horizons would tend to amplify the effects observed in Section \ref{sec:results}.

In the previous section, we assumed \textit{stagewise independence}. That is, the distribution of $\xi_{t+1}$ is independent of the distribution of $\xi_t$. This assumption is crucial for model tractability as it allows to approximate a single function of \textit{expected} cost-to-go. If, for example, a particularly windy stage $t$ were to make it more likely to encounter a particularly windy state in $t+1$, the problem would grow substantially in size or require additional simplifications \cite{lohndorf_modeling_2019}. 

We devise a simple test for the stagewise independence assumption. As detailed in Section \ref{sec:si_stagewise}, we compute the autocorrelation function for the deviations from the stage mean of all the input parameters entering $\xi$. Figure \ref{fig:acf_monthly} shows the autocorrelation for each input parameter for twelve monthly lags. Most importantly, solar PV, wind capacity factors and heat demand show no significant autocorrelation. Heat pump coefficients show a weakly significant first lag. By contrast, hydrological input factors for run-of-river, reservoirs and open pumped hydro storage inflows show significant autocorrelation. In the case study below, we exclude reservoirs and open pumped hydro to concentrate on hydrogen cavern LDES. While this simple test possibly neglects higher-order or cross-dependencies, we conclude that stagewise independence seems like a reasonable assumption for a monthly stage length. By contrast, we show in Section \ref{sec:si_stagewise} that weekly stages exhibit some first-order autocorrelation even for solar PV and wind capacity factors.

\subsection{Storage target level}\label{sec:storage_target}

The Limited Foresight model has twelve monthly subproblems, representing one operational year. CEMs are commonly used to find cost-optimal long-run equilibria. Without any additional restriction, the system planner has no incentive to retain energy in LDES beyond the final stage of a problem formulation with finite stages and would empty the storage greedily. This would be at odds with the notion of a \textit{steady-state} long-run equilibrium, in which the operational year is a representation of many. In deterministic settings, this is usually ensured by introducing a circularity constraint which connects the last time step to the first in the storage balance. In the stochastic setting considered here, this cannot be guaranteed. 

Instead, we introduce an additional state variable for long-duration storage $s$, $e_s^{ini}$. The system planner chooses the initial storage level along with generation and storage capacities in stage $0$, which determines the starting endowment at the start of stage $1$. In period $T$, any deviation of the ending storage level from $e_s^{ini}$ is penalized at a very high price (value of lost load). The system planner faces the trade-off of setting a high starting endowment at the risk of paying penalties if the storage target level is not met by the end of the model horizon. 

Such storage targets are not far-fetched. Similar targets have been set for natural gas storage facilities by the European Commission for several years now \cite{european_commission_commission_2024}. However, as we will see below, it represents a coarse approximation of the marginal value of storage beyond the last stage, which may bias dispatch and capacity decisions in an \textit{end-of-horizon effect} \cite{shapiro_periodical_2020}. 

Hole et al. \cite{hole_capacity_2025} recently used a discounted infinite horizon to eliminate such end-of-horizon effects in a capacity expansion model. The length of forward and backward passes of the SDDP algorithm becomes random \cite{dowson_policy_2020}. At reasonable discount rates, which determine the expected pass lengths, the number of subproblems to solve grows significantly. Computational tractability is maintained by a very reduced representation of dispatch operations. 

In this work, we opt to preserve a granular chronological sequence within each stage to capture interactions between LDES and other technologies at the expense of a potential \textit{end-of-horizon} effect.




\subsection{Long-duration storage bidding behavior} \label{sec:bidding_theory}

The trained Limited Foresight model consists of capacity choices in stage $0$ and the learned \textit{expected} cost-to-go functions that guide the dispatch trade-off between the current and future stages. By design, the current stage's dispatch decisions only have consequences for future stages in that they affect the LDES level through \textit{expected} opportunity costs. The hydroelectric literature refers to these costs as \textit{water} value or marginal storage value (MSV) and has long used stochastic programming methods to determine optimal hydro-storage bidding \cite{fleten_stochastic_2007,steeger_optimal_2014}. 

We derive the LDES bidding behavior in the Limited foresight model, building on Brown et al. \cite{brown_price_2025}.
Each dispatch stage is a linear program, for which the Karush-Kuhn-Tucker (KKT) conditions give necessary and sufficient optimality conditions.

In the optimum of stage $t$, the KKT stationarity condition with respect to LDES dispatch is given by,
\begin{subequations}
\begin{align}\label{eq:kkt_discharge}
            0 = \lambda_{t,h} - (\eta^f_s)^{-1}\lambda_{t,h}^{MSV} + \underline{\mu}^f_{s,t,h} - \bar{\mu}^f_{s,t,h} \quad \forall h \in \mathcal{H}(t),s
\end{align}    

\noindent where $\lambda_{t,h}$ is the electricity price in period $h$, $\lambda_{s,t,h}^{MSV}$ is the MSV of LDES $s$, $\eta_s^f$ is the discharge efficiency and $\underline{\mu}_{s,t,h}^f$, and $\bar{\mu}_{s,t,h}^f$ are the dual variables to the hydrogen turbine (LDES dispatch) capacity bounds $[0,H_s]$. Likewise, the storage charging condition is given by,
    \begin{equation} \label{eq:kkt_charge}
        \begin{aligned}
            0 = \eta^h_s\lambda_{t,h}^{MSV} - \lambda_{t,h} + \underline{\mu}^h_{s,t,h} - \bar{\mu}^h_{s,t,h} \quad \forall h \in \mathcal{H}(t),s
        \end{aligned}
    \end{equation}

When LDES is at the margin the market price for electricity is given by either $\lambda_{t,h} = (\eta^f_s)^{-1}\lambda_{t,s,h}^{MSV}$ or $\lambda_{t,h} = \eta^h_s\lambda_{s,t,h}^{MSV}$. Hence, a storage bids the efficiency-adjusted MSV. Assuming a hydrogen storage with $\eta^f = 40\%$ (hydrogen turbine) and $\eta^h = 70\%$ (PEM electrolysis), and a marginal value of storage of 100 € \ MWh$^{-1}_{H2}$, the LDES would bid at 250 € \ MWh$^{-1}_{el}$ to discharge and at 70 € \ MWh$^{-1}_{el}$ to charge. The MSV itself is defined by the stationarity condition with respect to the storage level $e_{s,t,h}$,
\begin{equation} \label{eq:msv_opp_1}
    \begin{aligned}
        0 = \lambda^{MSV}_{s,t,h} -\lambda^{MSV}_{s,t,h+1} + \underline{\mu}_{s,t,h}^{e} - \bar{\mu}_{s,t,h}^{e} \quad \forall h \in \mathcal{H}(t)\setminus \{H\},s
    \end{aligned}
\end{equation}
\noindent where $\underline{\mu}_{s,t,h}^e$ and $\bar{\mu}_{s,t,h}^e$ are the dual variables to the storage level bounds $[0,E_s]$. In the last period $H(t)$ of each stage, the condition is given by,
\begin{equation}\label{eq:msv_opp_2}
    \begin{aligned}
         \lambda^{MSV}_{s,t,H} = \frac{\partial\mathcal{Q}_{t+1}(\cdot,e_{s,t,H})}{\partial e_{s,t,H}} + \underline{\mu}_{s,t,H}^{e} - \bar{\mu}_{s,t,H}^{e} \quad \forall s
    \end{aligned}
\end{equation}

The partial (sub-)derivative of the \textit{expected} cost-to-go with respect to the LDES ending level \emph{at} the current ending level is simply the \textit{expected} MSV at the beginning of the next stage, conditional on the LDES ending level $e_{s,t,H}$. We also clarify that the expectation is taken with respect to the next stage's weather vector $\xi_{t+1}$ and can express the MSV in stage $t$ and period $h$ as,
\begin{align}\label{eq:nested_bidding}
    \lambda^{MSV}_{s,t,h} = \sum_{j=h}^{H(t)}(\underline{\mu}_{s,t,j}^e - \bar{\mu}_{s,t,j}^e) + \mathbb{E}\left[ \lambda_{s,t+1,1}^{MSV}(e_{s,t,H},\xi_{t+1}) \right]
\end{align}
\end{subequations}
This recursive structure is at the heart of LDES bidding under uncertainty. The MSV reflects the consequences of all possible weather realizations from $t$ onward. If in a stage $t'>t$ under a particular weather trajectory, the storage runs empty in period $h$, $\underline{\mu}_{s,t',h}$ turns positive, reflecting the cost of an outside option, e.g. load-shedding or energy imports. Such events in specific weather realizations from $t+1$ onward feed back to $t$ through the nested tree of possible weather realizations and enter the expectation in the expression for $\lambda_{s,t,h}^{MSV}$. Thereby, the MSV represents the \textit{conditional} probability of a bad system state times the marginal costs associated with such a state. The probability is conditional on the current storage level and on the remaining dispatch horizon. A higher storage level reduces the probability of bad states occurring in the future and the MSV goes down. Likewise, having a low storage level in November with imminent \textit{Dunkelflaute} periods leads to higher MSVs than having the same storage level in March, when the worst of winter is over and the system operator can expect sufficient VRE supply in the coming months.



SDDP builds outer approximations of the piece-wise linear $\mathcal{Q}-$functions, and the trained model can be used to back out the marginal storage values as functions of the current month (stage) and storage level. We do so by evaluating the partial (sub-)derivatives (slope) of $\{\hat{\mathcal{Q}}_t\}_{t=1}^T$ with respect to the stage's LDES ending level over the domain of the storage level $[0,E_s]$ in 10 GWh steps. We refer to these functions as LDES bidding curves.

For the last month $T$, the expected MSV is equal to the penalty for any ending level below the initial storage level $e^{ini}_s$ and zero above.

\subsection{Representation of risk preferences and outside options}

In the context of interannual weather variability, the aim of CEMs is often to find \textit{resilient} or \textit{robust} system designs that avoid load shedding or the extensive use of an expensive backup technology. In a deterministic setting, one can simply exclude the possibility of energy balance violations or attach a sufficiently high price for them not to occur. The stochastic setting considered here requires a more careful consideration of societal risk preferences. 

The Limited Foresight model above minimizes \textit{expected} system costs. The implicit assumption is that market participants -- through the agency of a system operator -- are risk-neutral towards weather risk \cite{shapiro_lectures_2009,roald_power_2023}. This may not be very realistic. An obvious alternative would be a risk-averse reformulation of the capacity expansion model \cite{munoz_does_2017,roald_power_2023}, which is generally possible for SDDP models \cite{shapiro_risk_2013}. Under the (strong) assumption of complete financial markets, there exists a system operator risk-averse model equivalent to the market outcome of risk-averse participants \cite{ehrenmann_generation_2011,dimanchev_consequences_2024}. However, it is not clear how to determine the right representation and degree of risk aversion. 

Coherent risk measures, like conditional value-at-risk, have a dual representation \cite{shapiro_lectures_2009}, which is equivalent to an optimization of an expectation over a \textit{changed} probability distribution \cite{liu_risk_2020} that puts more weight on adverse outcomes. A more risk-averse system operator would attach a higher probability to those scenarios in which the storage runs empty and the LDES dispatch would be more conservative. It should be noted that increasing the costs of bad system states instead of attaching a higher probability has a similar effect in that it increases the expected marginal value of storage. Without further formalization, we simply take away that varying the availability and costs of backup options may have a secondary interpretation of varying the degree of risk aversion the system operator exerts when operating the LDES. 

In the case study below, we vary the availability of outside options. The base scenario assumes a very high value of lost load and no alternative backup to enable a direct comparison of the limited foresight model with a perfect foresight model that aims to be resilient to weather variability.




\section{Case study}\label{sec:case_study}

\begin{figure}
    \centering
    \includegraphics[width=0.9\linewidth]{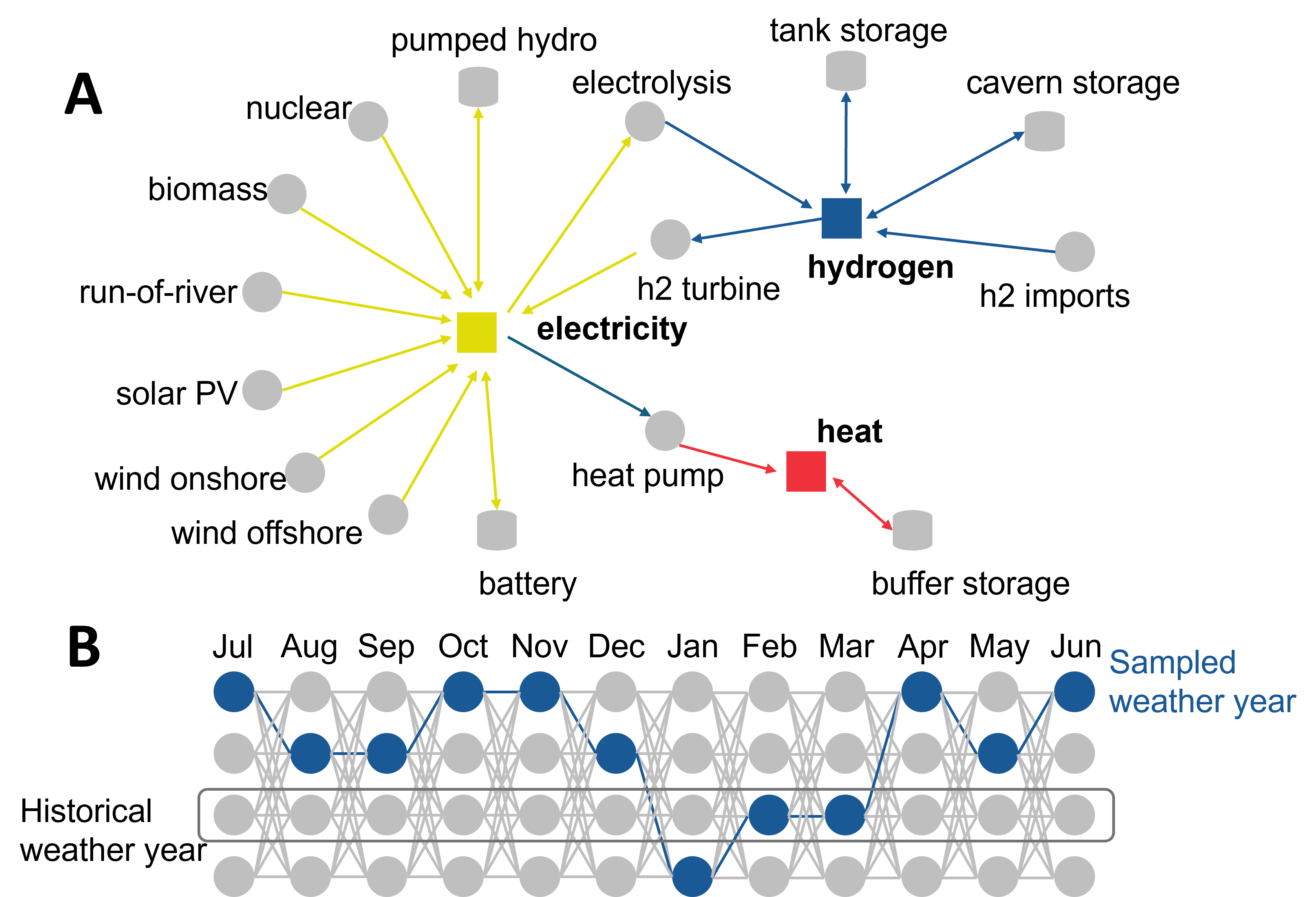}
    \caption{\textit{System representation and sampling lattice}: Panel A shows a stylized representation of the considered system. Electricity generation is confined to renewable sources. In the case of Spain and the UK, existing nuclear capacities are also considered. Electrified heat demand is served by a single type of heat pump. Hydrogen is produced via electrolysis or imported by ship, either under long-term contracts or, depending on the scenario, via spot shipments. Hydrogen can be stored in pressurized steel tanks or caverns and combusted in turbines to generate electricity. Panel B is a partial representation of the scenario lattice representing sample space $\Xi$. Each node represents a historical weather month. A sample path $\xi_i$ can be any combination of monthly nodes, like the example trajectory of darkblue nodes.}
    \label{fig:case_study}
\end{figure}

We isolate the effect of limiting foresight on LDES bidding behavior and optimal capacity choices by comparing a capacity expansion model of type D (Limited Foresight, LF) to a capacity expansion model of type C (Perfect Foresight, PF).

\subsection{Model description}

For both models, we use an adapted version of the open-source dispatch and capacity expansion model \textit{Dispatch and Investment Evaluation Tool with Endogenous Renewables} (DIETER) \cite{zerrahn_long-run_2017,schill_long-run_2018}. The model has been used in various studies to investigate flexibility in sector-coupled VRE-based energy systems \cite[e.g.][]{zerrahn_economics_2018,schill_electricity_2020,roth_geographical_2023,roth_power_2024,gueret_impacts_2024}. DIETER is a linear cost-minimizing model that abstracts from discrete capacity decisions and economies of scale as well as non-convex operational constraints. The model is centered around the power sector with coupled hydrogen and heating sectors, see Figure \ref{fig:case_study}, panel A. We describe a simplified formal model in Section \ref{sec:si_model_description}.

The model version used in this case study strikes a balance between capturing most of the relevant dynamics in a sector-coupled decarbonized energy system and maintaining computational tractability. As Section \ref{sec:convergence_details} shows, the SDDP algorithm solves millions of monthly subproblems in order to identify an approximately optimal policy. We reduce the computational burden by averaging all time series to four-hourly time steps, which tends to have negligible effects on optimal capacities \cite{schlachtberger_cost_2018}. We solve separate problems for three single-node systems. Germany, Spain and the United Kingdom (UK) 
represent a balanced, a solar-dominated and a wind-dominated system respectively. These countries have also been used in other studies concentrating on LDES and weather variability \cite{brown_ultra-long-duration_2023}. We abstract from any electricity or hydrogen exchange with neighboring countries, save for hydrogen imports at exogenous prices. The modeled target year is 2050, and only zero-emission generation technologies are admissible candidates in a static, \textit{one-shot} capacity investment stage. 
 
\subsection{Input data} \label{sec:input}

The candidate generation portfolio is comprised of solar PV, on- and offshore wind, bioenergy, run-of-river hydroelectric plants and nuclear capacity for the case of Spain and the UK. Storage technologies in the power sector include lithium-ion batteries and closed pumped-hydro storage. We assume that these short-term storages are subject to a circularity constraint within each month. We abstract from other hydro-storage technologies with inflow.\footnote{Reservoirs would require a similar treatment as LDES. Their inclusion would consequently increase the state space of the SDDP model, as their storage level would also need to be tracked over the months. The stagewise independence test in Section \ref{sec:stagewise} additionally suggests that there is some autoregressive structure in hydro inflows, which would further complicate an adequate representation.}

For the hydrogen (H2) sector, the model can invest in polymer electrolyte membrane (PEM) electrolysis, underground H2 storage in salt caverns and pressurized tank storage above ground as well as hydrogen gas turbines for reconversion to electricity. Additionally, the system operator can enter into long-term contracts (LTC) for H2 that deliver a constant amount in every time step, subject to limited off-take flexibility of +/-10\%. 

For heat supply, we only consider air-sourced heat pumps with a small buffer storage of 1.5 hours. The heat pump output capacity is set exogenously to meet peak demand across all historical weather years in the sample.  

Almost all cost and technical parameters are taken from the Danish Energy Agency projections for 2050 \cite{danish_energy_agency_technology_2024}. We provide more details in Table \ref{tab:tech_data}. Technical capacity bounds for solar PV and wind onshore are taken from Tröndle et al. \cite{trondle_home-made_2019}. Wind offshore is constrained by the upper limit in the Ten Year Network Development
Plan (TYNDP) 2024 of the European Network of Transmission System Operators for electricity and gas (ENTSO-E and ENTSO-G) \cite{entso-e_and_entso-g_tyndp_2024} to account for other build-out constraints than just technical availability. As a lower bound for solar PV and wind we take the currently installed capacities. Biomass, run-of-river and nuclear generation capacities are constrained to National Estimates for 2030 in the European Resource Adequacy Assessment (ERAA) 2021 \cite{entso-e_european_2021}. We provide details in Tables \ref{tab:tech_bounds} and \ref{tab:sto_bounds}. Cavern storage potentials are taken from Caglayan et al. \cite{caglayan_technical_2020}.


For weather data, we use country-level time series for solar PV, wind and run-of-river capacity factors from the Pan-European Climate Database \cite{de_felice_entso-e_2022} and heat demand and heat pump coefficients of performance (COP) data from Göke \cite{goke_pecd_2024}. We obtain hourly time series for 1982-2017. We average the data to four-hourly periods and define summer-to-summer weather years from July to June. For each calendar month $t$, we compile a sample space $\Xi_t$ that contains 35 historical samples from that month, i.e. $\Xi_1$ contains data for 35 Julys from 1982/83 to 2016/17, $\Xi_2$ contains 35 Augusts and so on. Each sample comprises 168 to 186 time steps. This way, the sampling process preserves the inherent seasonality of the weather data. Figure \ref{fig:case_study}, panel B, illustrates the stagewise independent sampling process. Any combination of stratified weather months is admissible, resulting in a total sampling space $\Xi=\Xi_1\times\cdots\times\Xi_{12}$ of $|\Xi|=35^{12}\approx3.38\times10^{18}$ possible \textit{synthetic} weather years.

\subsection{Scenarios}

The comparison between Limited Foresight and Perfect Foresight is conducted for different countries and for different assumptions with respect to the availability of backup technologies and supply.


We have shown in Section \ref{sec:bidding_theory} that the cost of the outside option is a key input to the LDES bidding behavior. For that reason, we devise three different scenarios. The first scenario, \textsc{No Imports}, entails no backup supply but requires load shedding in case the LDES cannot serve positive residual load in the system. Load shedding is penalized at the value of lost load (VOLL), which we set very high at 100,000 € MWh$_{el}^{-1}$ \cite{gotske_designing_2024}. In a second scenario, \textsc{Constrained Imports}, we assume that the system operator can buy up to 5.5 GWh$_{H2}$ per hour of shipped H2 on a global spot market at a price of 250 € MWh$_{H2}$$^{-1}$, which corresponds to one ship of ammonia arriving at a time.\footnote{This rough constraint results from the following back-of-the-envelope calculation. One ship carries 60,000 tonnes of ammonia with a hydrogen energy content of 5.86 MWh$_{H2}$ ton$^{-1}$. After accounting for energy demand for cracking at 1.41 MWh$_{H2}$ ton$^{-1}$ and assuming an unloading time of 48 hours, we arrive at 5.5 GWh$_{H2}$ per hour.} If the constraint is binding, load shedding at VOLL is the last resort. The last scenario, \textsc{Unconstrained Imports}, allows unlimited imports of H2 from spot markets at 250 € MWh$_{H2}$$^{-1}$. We summarize these assumptions in Table \ref{tab:scenarios}.

\begin{table}[H]
    \centering
    \begin{tabular}{rrr}
    \hline
         Scenario& Cost of outside option & Maximum quantity\\ \hline 
         No Imports& 100,000 EUR MWh$_{el}^{-1}$ & $\infty$  \\
         Constrained Imports & 250 EUR MWh$_{H2}^{-1}$ & 5.5 GWh$_{H2}/h$ \\
         Unlimited Imports & 250 EUR MWh$_{H2}^{-1}$ & $\infty$ \\ \hline
    \end{tabular}
    \caption{\textit{Scenarios }}
    \label{tab:scenarios}
\end{table}

\subsection{Implementation} \label{sec:implementation}

We use an open-source ~\textsc{Julia}~ implementation of DIETER, DIETER.jl, and couple it with SDDP.jl \cite{dowson_sddpjl_2021}. 

\noindent We solve the Limited Foresight model for each country and scenario using the SDDP algorithm. Each model has 20 state variables representing generation and storage capacities, LDES levels and H2 import contracts. SDDP.jl features a multi-threading option that allows solving subproblems in parallel. We use four threads. Gurobi 10.0.1 is used to solve the linear subproblems with dual simplex. We set a time limit of 54 hours and an iteration limit of 15000 iterations. All computations are performed on a high-performance computing cluster node with a total memory of 256 GB. In order to prevent excessive oscillation of the capacity variables in the first stage, we use a simple trust-region stabilization \cite{goke_stabilized_2024}.

While SDDP has been shown to converge almost surely in finitely many iterations \cite{philpott_convergence_2008}, practical termination criteria remain an open research topic \cite{dowson_policy_2020}, as the upper bound is probabilistic. We opt for the aforementioned time and iteration limits and inspect convergence ex-post. In the Section \ref{sec:convergence_details}, we show that capacity decisions begin to stabilize after around 3000 iterations, see Figure \ref{fig:convergence}. Nevertheless, subsequent iterations serve to refine the policy, which is particularly useful in order to obtain granular LDES bidding curves. We report additional details on convergence for each model in Table \ref{tab:computational_res}. We facilitate a comparison to the Perfect Foresight model by simulating the 35 historical weather years $\{1982/83,\dots,2016/17\}$ in our sample using the trained policy $\{\hat{\mathcal{Q}}\}$. That is, we solve the dispatch for a given weather year by solving each monthly subproblem with the learned \textit{expected} cost-to-go functions. 

The Perfect Foresight model is a two-stage version of the CEM where the second stage has 2190 four-hourly periods. We also implement it in SDDP.jl for consistency but solve the deterministic equivalent linear program that includes all 35 weather years in the sample.

\section{Results} \label{sec:results}

\subsection{LDES stockpiles against bad system states} \label{sec:results_ldes_traj}

\begin{figure}[H]
    \centering
    \includegraphics[width=\linewidth]{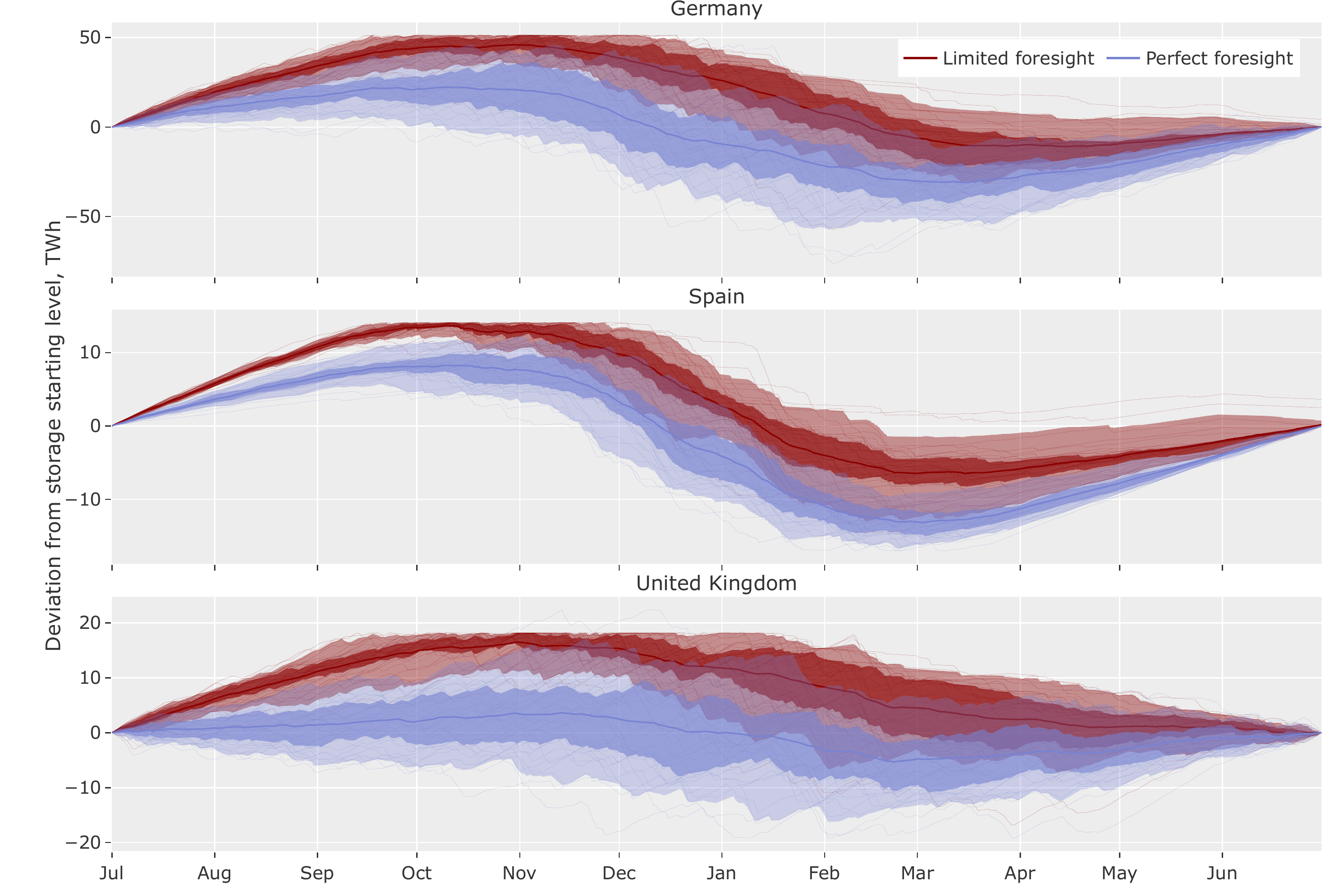}
    
    \caption{\textit{LDES storage level trajectories}: Panels show the distribution of LDES storage level trajectories over 35 historical years (July 1982-June 2017) for Germany (top), Spain (middle) and the United Kingdom (bottom). Red components refer to Limited Foresight, and blue components refer to Perfect Foresight. Trajectories are normalized by the endogenous storage target level. The thick lines show the mean trajectory. Darker shaded areas represent the 25th-75th percentiles, lighter areas show the 5th-95th percentile of the distribution. Thin lines show the actual trajectories.}
    \label{fig:ldes_traj}
\end{figure}

Fig. \ref{fig:ldes_traj} shows the distribution (5th-95th, 25th-75th percentile and mean trajectories as well as the actual paths) of storage trajectories when simulating the Perfect Foresight case (blue; PF) and the Limited Foresight case (red; LF) over the 35 historical weather years between July 1982 and June 2017. These results are for the \textit{No Imports} scenario. Fig. \ref{fig:traj_constr} and Fig. \ref{fig:traj_unlimit}  present the trajectories for the other two scenarios.

In both models, LF and PF, the storage needs to meet an endogenously chosen storage target level by the end of June. As these target levels and storage capacities may differ between PF and LF, we normalize the storage trajectories around the target level. This isolates the differences in storage behavior under each model specification. 

For all three countries, the mean trajectory in the LF case lies well above that of PF. This difference is largest in December when \textsc{LF} LDES levels are still close to full capacity, whereas PF LDES levels cross the zero line, entering into a \textit{deficit} vis-à-vis the target level.  
The LF model reaches mean levels close to full capacity as early as October and maintains them until the turn of the year in the case of the United Kingdom. This way, the LDES operator \textit{hedges} against bad weather states with prolonged periods of VRE scarcity potentially coinciding with cold spells \cite{schmidt_mix_2025}. In the \textit{No Imports} scenario, such states could become very costly, as load would have to be shed at the cost of VOLL if the LDES cannot supply during a positive net load event. By contrast, the PF model can foresee all periods of scarcity and abundance over the model horizon and behaves much more aggressively. Under PF, the maximum dispatch event across all weather years defines the storage size \cite{kittel_coping_2025} and only then full capacity is met.

Apart from these common dynamics across countries, there are also significant differences. The Spanish trajectories under LF are fairly close to those under PF. The LF mean trajectory enters the \textit{deficit} roughly one month after its PF counterpart but runs a negative balance of up to 6.5 TWh before returning to the target level. Both distributions, PF and LF, are rather narrow. The United Kingdom presents a different picture. The LF mean trajectory never enters a deficit. The maximum deficit across all years is just 3.9 TWh compared to 20 TWh under PF. Both distributions are rather wide. The PF distribution begins running wide much earlier than the LF distribution.

We present an alternative perspective in Figure \ref{fig:chrono_traj}, which gives the storage trajectories in their chronological order over the entire period from 1982 to 2017 and at absolute scale. Spain's trajectories are far more even, and the difference between LF and PF is rather stable over the sample, with the LF LDES consistently reaching its low around 10 TWh and the PF LDES frequently almost emptying the storage. By contrast, the United Kingdom's trajectory is far more erratic. While the storage level fluctuates wildly over the years in the PF case, it consistently charges to maximum capacity in preparation for the winter under LF. Interestingly, the PF storage level exceeds the LF storage level in few instances for Germany and the UK, enabling deeper dispatch episodes, especially around the \textit{Dunkelflaute} of 1996/97 and the cold winter of 2010/11 in the United Kingdom.

Spain and the United Kingdom represent solar-dominated and wind-dominated systems, respectively. As evident from Fig. \ref{fig:input_dist}, the empirical distribution of wind capacity factors is wide and wind tends to follow synoptic patterns \cite{schlott_impact_2018}. By contrast, seasonal and diurnal patterns for solar generation are much more stable, and the distribution tends to be narrow in comparison. The width of the PF distribution for the UK reflects sporadic dispatch in windless periods and charging in windy periods, which also occur in the winter period. Lacking perfect foresight, the LF model cannot rely on windy winter periods and stockpiles in expectation of extreme winter periods. In Spain, the seasonality is much more pronounced, which reflects predictably lower solar capacity factors in winter. Germany, neither wind- nor solar-dominated to the same degree, exhibits both patterns.

As the costs of bad system states reduce as hydrogen spot imports become available in scenarios \textit{Constrained Imports} and \textit{Unlimited Imports}, the differences between the two models dissipate as the optimal policy replaces some stockpiling by spot imports, see Figure \ref{fig:traj_constr} and Figure \ref{fig:traj_unlimit}.

\subsection{System value of solar PV increases under uncertainty} \label{sec:results_capacity}

\begin{figure}[H]
    \centering
    \includegraphics[width=\linewidth]{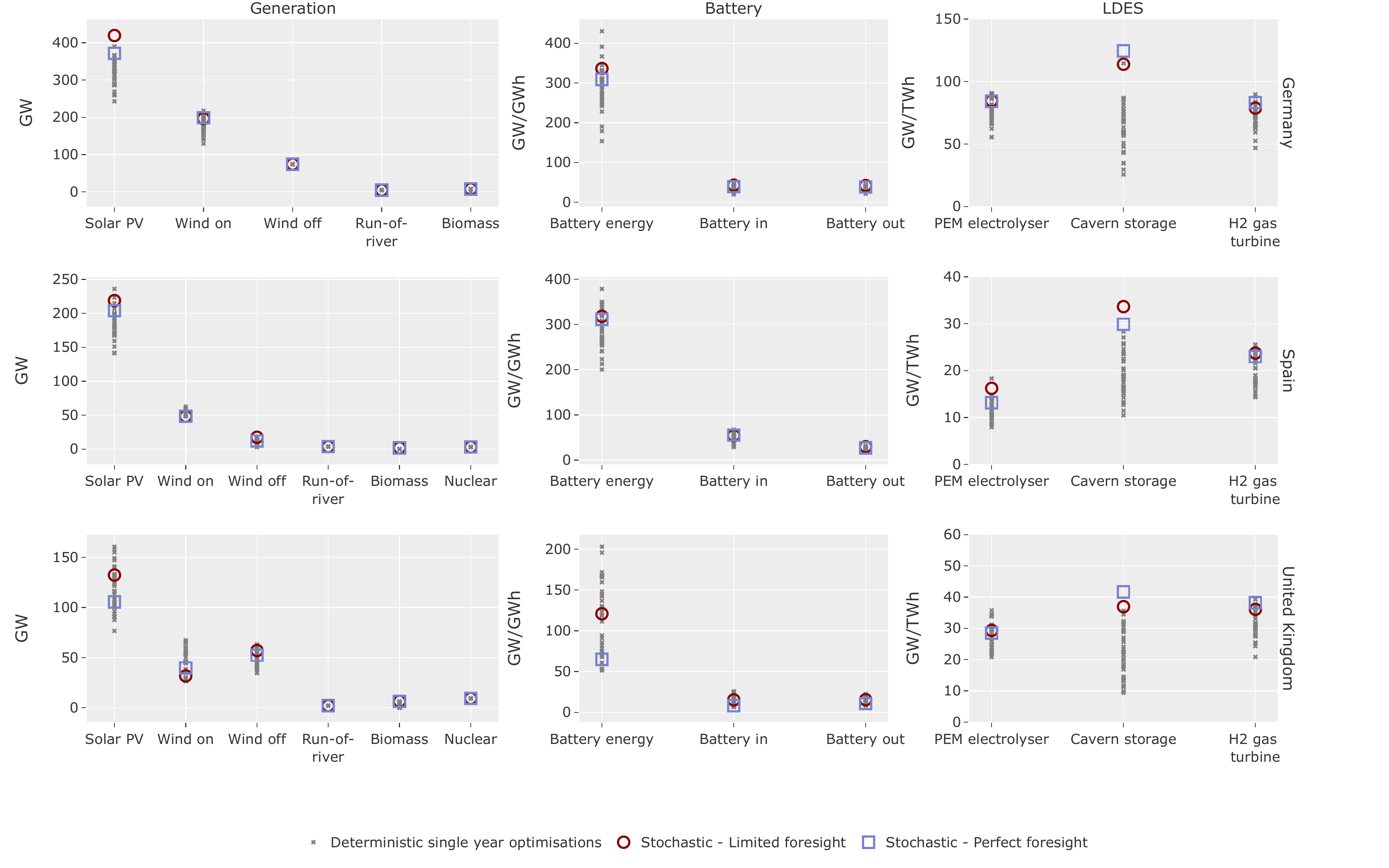}
    \caption{\textit{Comparison of capacity choices}: Assuming \textit{No Imports}; Left panels show generation capacities; middle panels show Li-Ion capacities; right panels show capacities for LDES components. In addition to Limited Foresight (red) and Perfect Foresight (blue), gray crosses represent deterministic single-year results for comparison.}
    \label{fig:capas}
\end{figure}

The observed differences in LDES behavior affect optimal capacity decisions. Again, we focus on the results in the \textit{No Imports} scenario. 
As Figure \ref{fig:capas} shows, the differences between PF (blue squares) and LF (red circles) are modest compared to the wide range observed when optimizing capacities for each weather year separately (type A from Section \ref{sec:theory}), as indicated by the gray crosses. The latter is especially wide for cavern storage capacity, reflecting LDES' sensitivity to interannual weather variability.

In terms of generation capacities, we observe a significant increase in solar PV capacity between PF and LF. For Germany, the increase is 48 GW or 12.8\%. For the UK, the increase is ca. 30 GW or 25.4\%. By contrast, Spain's PV capacity only increases by 7.4\% or 15 GW.  Wind onshore capacity differences are smaller. Only for the UK, we observe a larger reduction of 7.9 GW or 20\%. Wind offshore capacities remain at the upper limit for Germany defined in Section \ref{sec:input}. For the UK, they increase slightly (9\%). Having limited offshore potential, Spain's capacities increase significantly on a relative basis, from 12 to 17 GW. Run-of-river, nuclear and biomass are either fixed by design or unchanged between the scenarios.

Battery power capacities remain largely unchanged between PF and LF for Germany and Spain. Energy capacities increase slightly (9\% for Germany, 3\% for Spain). For the UK, the energy capacity grows significantly, 56 GWh or 86\%, to match the duration of batteries in Germany of 8 hours.

The impact of limited foresight on LDES capacities is rather small. PEM electrolysis capacities remain at par with PF in Germany and the UK. In Spain, they increase slightly by 3 GW. In Germany and the UK, cavern storage capacities decrease (DE: -8\%, UK: - 11\%), while they increase for Spain, albeit on a lower level. 

The described capacity differences between PF and LF are consistent with the differences in LDES behavior discussed in the previous section. Stockpiling requires reliable electrolysis activity during the summer and autumn months, including the period leading up to the storage target, to ensure high storage levels for bad system states that are more likely to occur in winter. Solar capacity factors have a much lower variance than wind capacity factors (Figure \ref{fig:input_dist}) and therefore offer more reliable feed-in in the lead-up to winter.

While under perfect foresight, the system operator can capitalize on particularly windy periods during winter that make up for periods of extreme energy scarcity, the LF model cannot leverage these episodes to replenish LDES levels as they are highly uncertain. Accordingly, solar generation replaces some wind onshore generation (Figure \ref{fig:gen_diff}). Electrolysis activity more closely reflects the diurnal patterns of solar PV feed-in (Figure \ref{fig:hourly_hm} and Table \ref{tab:corr_mat}) and is much more even across weather years, as opposed to PF, where it follows year-specific wind patterns (Figure \ref{fig:ely_recon_hm}). 

The change in LDES capacities can be explained by the chronological storage trajectories in Figure \ref{fig:chrono_traj}. For Germany and the UK, PF can leverage infrequent periods of abundance to store more the storage prior to a large discharge due to a Dunkelflaute or cold spell. With perfect foresight, the benefit of the required extra energy capacity outweighs the cost. The same does not hold under LF, where many more weather realizations are possible. In Spain another effect, in which the LF model keeps some insurance for extreme states that do not materialize over the historical weather year sequence, dominates and the cavern size increases under LF. 

The change in battery capacity likely serves to the improve the full-load hours of electrolysis capacity by smoothing solar feed-in. Additionally, previous work has shown that under PF, LDES may also engage in shorter-term balancing in order to reduce battery capacity needs \cite{kittel_coping_2025}. Under LF, the need for stockpiling likely outweighs the benefit of reducing battery capacity by running existing LDES capacity for short-term balancing. 

Unsurprisingly, the described effects tend to be largest for a wind-dominated system like the UK and smallest for a solar-dominated system like Spain. With additional backup from H2 spot imports in \textit{Constrained Imports} and \textit{Unlimited Imports}, the differences between PF and LF converge in line with the dissipation of the stockpiling behavior discussed above.



\subsection{LDES bidding functions} \label{sec:results_ldes_bidding}

\begin{figure}[H]
    \centering
    \includegraphics[width=\linewidth]{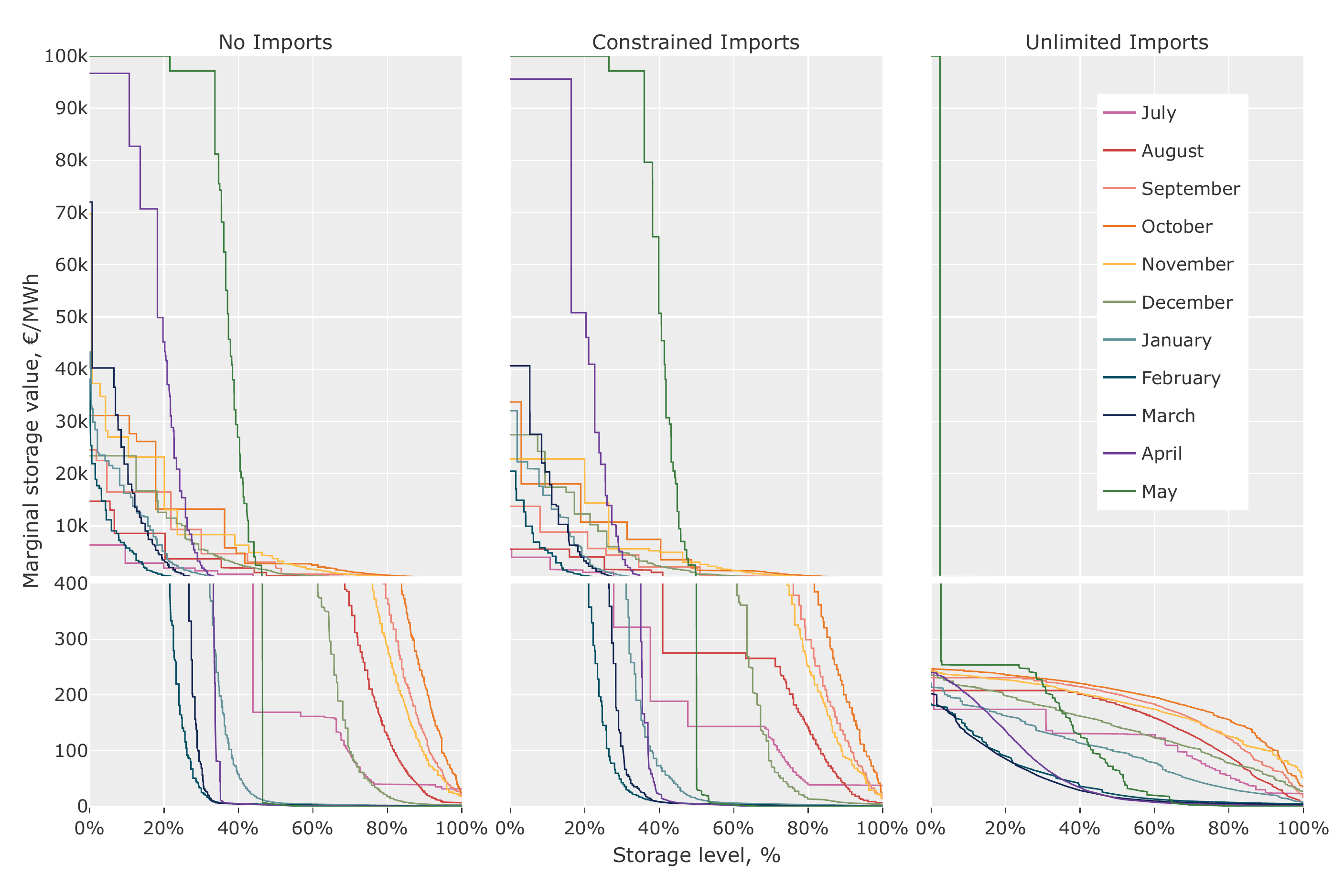}
    
    \caption{\textit{LDES bidding curves}: Monthly LDES marginal values of storage derived from trained SDDP model differentiated by alternative H2 import options for Germany.}
    
    \label{fig:bidding}
\end{figure}

Based on the derivations in Section \ref{sec:bidding_theory}, 
we compute LDES marginal storage values (MSV) or bidding functions. We evaluate the (sub-)derivative of each month's $\hat{\mathcal{Q}}$-function over the storage level domain $[0,E_s]$ in steps of 10 GWh. We show the resulting curves for Germany in Figure \ref{fig:bidding}. The y-axis contains a break at 400 € MWh$^{-1}_{H2}$ to accommodate the wide range of MSV.

Each point on a curve reflects the \textit{expected} MSV in a given month at a given storage level. For example, at a storage level of 80\% by the end of August, the German LDES would bid 127 € MWh$^{-1}_{H2}$ (Figure \ref{fig:bidding}, left panel). After accounting for electrolysis and reconversion efficiencies, this would correspond to a bid of 89 € MWh$^{-1}_{el}$ to consume electricity and an offer price of 318 € MWh$^{-1}_{el}$ to produce electricity. The (estimate of the) expectation encompasses all possible weather realizations from September onward, assuming that the system is dispatched optimally using the learned bidding curves for the following months and conditional on the current storage level being 80\%.

For most possible weather realizations, the 80\% storage level, and everything that will be optimally added in a given realization in the following months, will suffice to cover all positive net load events over the model horizon. In these realizations, the MSV will be low. In a few extreme cases, the storage level will not suffice to cover all scarcity periods. In these instances, the MSV is set by the VOLL or by H2 spot import prices. The LDES bid in August at 80\% reflects the average over these possible outcomes, and although the probability of realizations with load shedding may be low, the assumed VOLL is so high that it drives up the average MSV. As we consider lower storage levels in the same month, the probability of load shedding or H2 spot imports increases and the associated MSV increases too. 

Likewise, the probability of incurring a future bad state changes by month. Prior to the winter period, the probability of a future energy shortfall at a given storage level is higher than after the worst of winter is over. Consequently, the autumn curves (yellow, orange, red and pink) have a concave shape in which high bids prevail at very high storage levels. Conversely, spring curves (blue, purple, green) are convex with low bids at relatively high storage levels. 

Towards the end of the dispatch horizon, the storage target affects LDES bidding. If the LDES level falls short of the target by the end of June, a penalty of 100,000 € MWh$^{-1}_{H2}$ is incurred. Consequently, the MSV in June is 100,000 € MWh$^{-1}_{H2}$ for storage levels below the target and 0 € MWh$^{-1}_{H2}$ for storage levels above the target. In order to reduce complexity, we have omitted this trivial bidding curve from the figure. However, even in prior months, MSVs increase rapidly as a shortfall against the storage target becomes more likely. This reflects the end-of-horizon effect discussed in Section \ref{sec:storage_target}. In an infinite-horizon setup \cite{hole_capacity_2025}, one would expect the curves to transition to the concave autumn curves more gradually. Similarly, the concavity of the autumn curves would be less pronounced, as positive MSV values above the storage target level would induce the stockpiling to commence earlier.

A comparison between the three scenarios explains the high influence of the outside option on LDES behavior. In the extreme case of \textit{Unlimited Imports}, the MSV is bounded by the price of imports at 250 € MWh$^{-1}_{H2}$ for most months.\footnote{Only in case of very low storage levels in May, there are not enough LDES power capacity to accommodate enough imports to meet the target and the MSV curve shoots up to 100,000 € MWh$^{-1}_{H2}$.} 

By contrast, the \textit{No Imports} scenario leads to extremely high bids for low storage levels in most months. Pre-winter months show bids up to 40,000 € MWh$^{-1}_{H2}$ for an empty storage. They only fall below 400 € MWh$^{-1}_{H2}$ above the 80\% level. This explains the stockpiling behavior and the implicit capacity effects observed in previous sections. 

The \textit{Constrained Imports} case suggests that the assumed spot import capacity is not sufficient to dramatically reduce MSVs across the curves compared to the \textit{No Imports} case.

In summary, it is the probability of extreme weather realizations in which the LDES cannot cover all the load and the associated costs to deal with them that determines the LDES bid. The probability changes with the time of year and the storage level. Naturally, it depends on the joint distribution of weather variables and the optimal dispatch given a weather realization, which are unknown (ex-ante). Using SDDP, the LF model \textit{learns} extreme state probabilities reflected in the bidding functions. The bidding functions become coarser (larger step sizes) towards low storage levels, as the model is unlikely to visit this state often in the forward pass simulations.

\subsection{LDES Bids under uncertainty stabilize energy-only market prices} \label{sec:price_stability}

\begin{figure}[H]
    \centering
    \includegraphics[width=\linewidth]{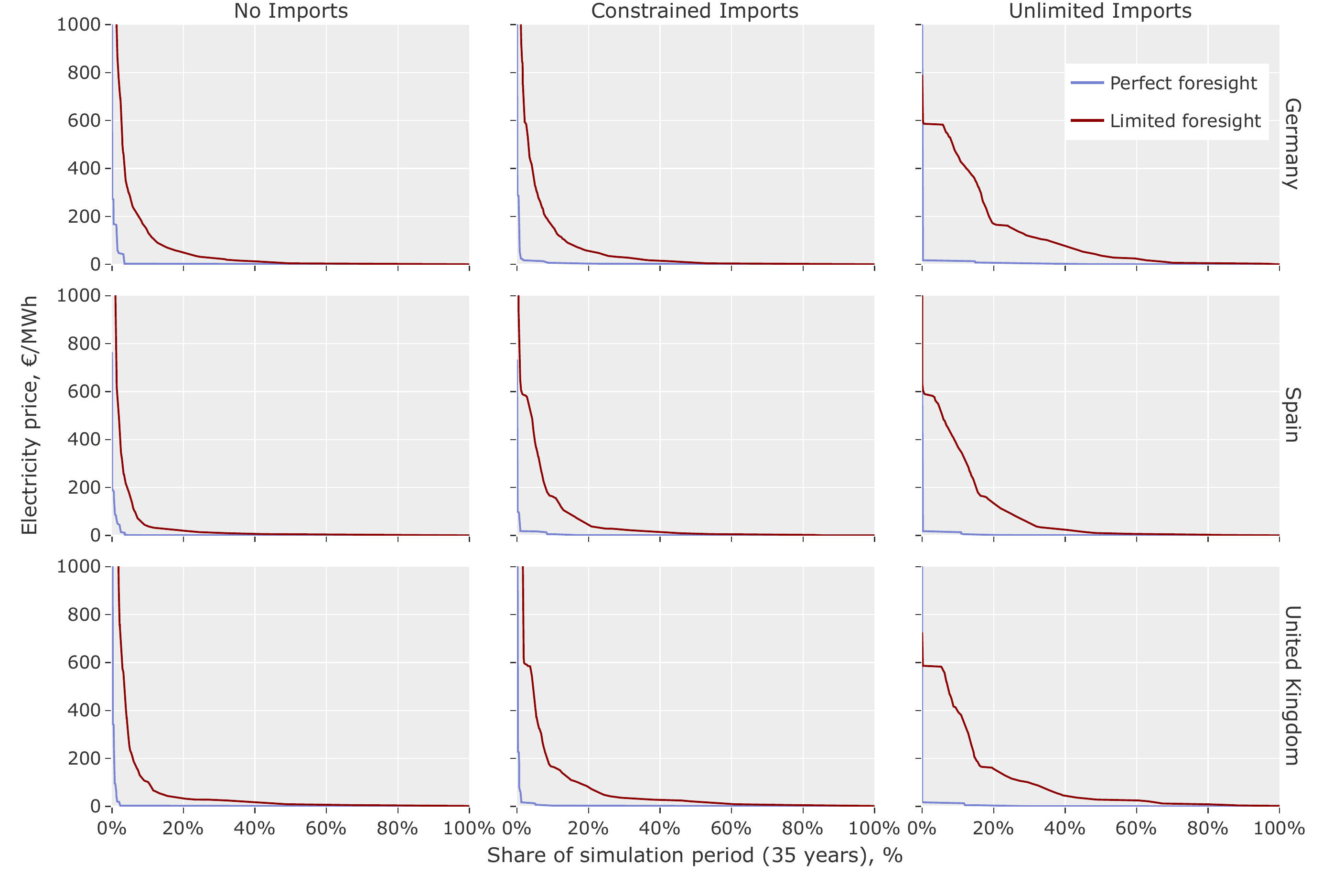}
    \caption{\textit{Electricity price duration curves over 35 simulated weather years}}
    \label{fig:duration}
\end{figure}

Finally, we investigate the impact of limited foresight on electricity prices. Fig. \ref{fig:duration} shows the electricity price duration curves for the full 35-year simulation period for both PF and LF in each country and under each import scenario. The red lines, representing duration curves under LF, have considerably higher shares of elevated prices than the blue lines showing PF prices. The latter exhibit extremely high shares of zero or near-zero prices and few hours with very high prices exceeding the axis upper bound of 1000 € MWh$^{-1}_{el}$. 

The explicit representation of weather uncertainty and the resulting LDES bidding behavior, discussed in the previous section, yield a wide range of bids at which the LDES may set the price, either charging (electrolysis) or discharging (H2 turbine), effectively smoothing the duration curve. By contrast, the \textit{possibility} of a bad state is not priced into the perfect foresight LDES bids, reflected in electricity prices when LDES is at the margin. That is, the PF LDES bids do not include the value of storage against situations that never materialize, resulting in a low MSV variation and electricity price \textit{disparity} in the (close to) zero marginal cost systems. Figures \ref{fig:msv_realized_no_imports}, \ref{fig:msv_realized_constr_imports} and \ref{fig:msv_realized_unlimited} compare the realized marginal storage values of LF and PF. 

The stabilizing effect of limiting foresight is most pronounced in the unlimited H2 imports case, where the probability of triggering the comparatively low-cost backup option (including the H2 turbine efficiency, the marginal price is at 581 €/MWh) is much higher than under load-shedding and the storage traverses the bidding curve more widely leading to more varied realized bids.


\section{Discussion and conclusion} \label{sec:discussion}

\subsection{Summary}

In this work, we use a method developed for the optimization of hydro-dominated electricity systems to optimize a simplified capacity expansion model for fully renewable sector-coupled European energy systems based on wind and solar. We isolate the effect of limited foresight under weather uncertainty by comparing the output to a perfect foresight benchmark. Under weather uncertainty, LDES bidding is defined by the probability and cost of extreme system states requiring load shedding or some type of backup. The probability is conditional on the current storage level and the time of year. LDES bids therefore depend on the underlying joint weather distribution (of wind, solar, hydro and heat demand). 

The bidding behavior leads to stockpiling in which the LDES builds up high storage levels to insure against \textit{potential} bad system states. By contrast, the perfect foresight model can foresee all periods of abundance and scarcity and \textit{tailors} the LDES trajectory to the residual load time series. The differences are most pronounced for a wind-dominated system like the United Kingdom. We find fewer differences for solar-dominated Spain.

We observe indirect effects on capacity decisions. While there are only smaller changes in the LDES capacities (electrolysis, cavern storage, hydrogen gas turbine), solar PV gains in system value for providing more reliable generation than wind, especially for the production of electrolytic hydrogen in the stockpiling process. We find the highest effect in the United Kingdom, where solar PV capacity increases by 25\% and onshore wind capacity decreases by 20\% under limited foresight.

Lastly, we demonstrate that LDES bidding under uncertainty critically affects price formation in fully renewable energy-only electricity markets by smoothing price duration curves as price-setting LDES bids constantly change to reflect the changed probability of extreme system states. 

\subsection{Implications}
There is a broad literature consensus that the design of variable renewable energy systems requires taking into account interannual weather \textit{variability} \cite{pfenninger_dealing_2017,collins_impacts_2018,ruggles_planning_2024,gotske_designing_2024,goke_stabilized_2024,grochowicz_intersecting_2023,grochowicz_using_2024}, especially in the context of long-duration energy storage (LDES) serving as backup during prolonged periods of energy scarcity \cite{brown_ultra-long-duration_2023,dowling_role_2020}. However, the role of limited foresight with respect to weather uncertainty in capacity expansion models is often ignored. Few studies have considered limited foresight in LDES operations and have either done so by using heuristic bidding rules \cite{brown_price_2025,dupre_la_tour_towards_2023}, or by using very reduced scenario trees \cite{blanchard_strategic_2024,hummelen_exploring_2024} in partial capacity expansion models. 

Ignoring weather uncertainty, capacity expansion planning can lead to biased systems designs and operations. Perfect foresight models tend to rely too much on wind capacities, as they can perfectly leverage intermittent windy episodes throughout an operational year. In a limited foresight model, solar PV gains in system value as its capacity factors are less variable, aiding the LDES stockpiling process. Meanwhile, LDES capacities themselves do not change much, and energy capacities remain close to those defined by the largest dispatch events such as the winters of 1996/97 and 2010/11 under PF as found in previous literature contributions \cite{kittel_coping_2025,ruhnau_storage_2022}. 

It is obvious that highly resolved sector-coupled capacity expansion models would be intractable to solve with a stochastic (dual) dynamic programming method like we apply here. Yet, the derived LDES bidding curves may serve as a starting point to improve on existing heuristics for LDES bidding rules \cite{brown_price_2025,dupre_la_tour_towards_2023}. They should be built on the insight that conditional marginal storage value derives from the cost and probability of extreme system states, which in turn varies with the underlying seasonal weather distribution.

The introduction of limited foresight also has important implications for price formation. The energy crisis in the wake of the Russian invasion of Ukraine has reignited a debate on electricity market design. Price formation on electricity markets with very high shares of variable renewable energy has been called into question as a substantial price disparity with long periods of near-zero prices and sporadic extreme prices threatens market functioning, cost recovery and raises financing costs \cite{mallapragada_electricity_2022}. The prices in the perfect foresight model above resemble these patterns. However, once we introduce limited foresight and the long-duration storage bids a marginal storage value that represents the probability-adjusted costs of extreme system states, we observe much smoother price duration curves that may solve many of the aforementioned concerns. These findings are similar to those of \cite{aaslid_pricing_2021} in a hydroelectric context and complement recent contributions by Brown et al. \cite{brown_price_2025} and Antweiler and Muesgens \cite{antweiler_new_2025} who highlight the role of demand elasticity and storage.

\subsection{Limitations}

The presented analysis is aimed at illustrating the mechanism through which weather uncertainty and LDES storage operations affect renewable energy systems and giving a sense of the order of magnitude of the effects. Nevertheless, there are several limitations that need to be considered when interpreting the results.

First, the case study adopts a very simplified energy system representation to maintain computational tractability. Each model is confined to a single country-node. Assuming away all interconnection with other countries in the European network is likely to exaggerate LDES requirements of extreme system states \cite{roth_geographical_2023,kittel_coping_2025}. Also the representation of sector-coupling is overly simplified. For example, long-duration thermal storage in the heating sector may provide a significant amount of the flexibility provided by LDES in the model \cite{schmidt_mix_2025}. It has been shown that sector-coupling plays a smoothing role in price formation, which may mitigate the extreme price bifurcation seen in the perfect foresight model above \cite{brown_price_2025}. Likewise, we omit flexible hydro capacities.

Second, we only consider hydrogen spot imports as a backup to the system neglecting other options to avoid load shedding including conventional generators, which could deal with some extreme system states at the cost of greenhouse gas emissions. Even in the absence of a backup technology, the correct estimation of the value of lost load is crucial in deriving a reasonable LDES bidding behavior and has consequences for the optimal system design. Here we assume a very high VOLL of 100,000 € MWh$_{el}^{-1}$.

Third, we assume monthly stage lengths to strike a balance between the validity of the stagewise independence assumption and a realistic degree of foresight. The latter is likely to be overly optimistic. However, overestimating foresight probably results in a lower bound to the differences between perfect and limited foresight and we would expect effects to be stronger over shorter foresight horizons. A violation of the stagewise independence assumption leads to meteorologically implausible weather years that may lead to substantial biases.

Fourth, the results indicate that there is a significant end-of-horizon effect \cite{dowson_policy_2020}. The LDES does not \textit{see} any value in adding storage levels beyond the target level at 1st July. In a multi-year or infinite-horizon formulation, the next winter would induce the storage to start stockpiling again earlier. The seasonal differences between bidding curves would likely be less pronounced as bids come down for autumn months and come up for spring and early summer months.

\subsection{Future research}

This analysis is meant to give an insight into the consequences of LDES bidding under weather uncertainty. We see several avenues for future research:

\textit{First}, one could further investigate the role of different marginal weather distributions. For example, one could seek to isolate the effect of heat demand uncertainty on LDES bidding curves. \textit{Second}, the LDES bidding curves trained on historical wind and solar data could serve as a starting point to build more refined heuristic bidding rules like those suggested by Brown et al. and Dupre la Tour \cite{brown_price_2025,dupre_la_tour_towards_2023}, to emulate limited LDES foresight in large-scale capacity expansion models. Resulting non-linearities could be addressed with piece-wise linear approximations that have been frequently used in the context of technological learning \cite{behrens_reviewing_2024}. Another avenue could be to use linear decision rules to approximate multi-stage decision-making \cite{egging_linear_2017,nazare_solving_2023}. \textit{Third}, the role of LDES bidding under uncertainty for price formation in fully renewable systems should be investigated further. There are likely to be interactions with demand elasticity, sector-coupled demand-side flexibility, such as long-duration thermal storage in district heating networks \cite{schmidt_mix_2025}, and interconnection. It may have consequences for optimal market design in a system with only VRE and storage \cite{billimoria_contract_2023}.

\section*{Code and data availability}
\noindent The open-source code used in this analysis is available on Github at \href{https://github.com/FSchmidtDIW/DIETER-SDDP}{https://github.com/FSchmidtDIW/DIETER-SDDP}. All results data are available in a Zenodo repository at \href{https://zenodo.org/records/15570551}{https://zenodo.org/records/15570551}. All input data can be fully reproduced using the open-source repo \textit{DieterData} available in a \href{https://gitlab.com/diw-evu/dieter_public/dieterdata}{https://gitlab.com/diw-evu/dieter\_public/dieterdata}.

\section*{Acknowledgments}

\noindent We thank the entire research group "Transformation of the Energy Economy" at the German Institute for Economic Research (DIW Berlin) for valuable discussions and inputs. Likewise, we would like to thank Prof. Tom Brown and the Department of Digital Transformation in Energy Systems at Technical University Berlin and Leonard Göke for fruitful discussions as well as Oscar Dowson for being extremely helpful in the model implementation process using SDDP.jl. We acknowledge a research grant from the German Federal Ministry of Education and Research via the \textit{Ariadne} projects (FKZ 03SFK5NO-2).

\printbibliography

\appendix
\setcounter{figure}{0}
\renewcommand{\thefigure}{SI.\arabic{figure}}
\setcounter{table}{0}
\renewcommand{\thetable}{SI.\arabic{table}}
\renewcommand{\thesubsection}{SI.\arabic{subsection}}
\setcounter{equation}{0}
\renewcommand{\theequation}{SI.\arabic{equation}}

\newpage
\section*{Supplemental Information}

\subsection{Formal model description} \label{sec:si_model_description}

\subsubsection{The Limited Foresight model}

This section provides a formal description of the multi-stage stochastic program that represents the Limited Foresight model. For simplicity, we focus on the electricity sector and abstract from the heat balance and simply include electricity demand from heat in the electricity energy balance. That way, we also abstract from some short-term flexibility in the heating sector included in the case study model. We also do not show the hydrogen balance in which hydrogen supply enters by way of electrolysis or imports and hydrogen demand arises from exogenous industrial demand and hydrogen gas turbines. Instead, we present the long-duration energy storage consisting of electrolysis, caverns and H2 turbines as if it were a generic electricity storage technology and abstract from its integration in the hydrogen sector. Again, the case study of Section \ref{sec:case_study} is more detailed and represents the hydrogen sector explicitly, also accounting for electricity demand from compressors as described in Kittel et al. \cite{kittel_coping_2025}. 

The problem has $T+1$ stages, where stage $0$ is the capacity investment stage and $t=1,\dots,T$ are dispatch stages. In the case study, these are assumed to cover a month of four-hourly periods, each resulting in stage lengths of $H(t) = |\mathcal{H}(t)|$ between 168 and 186 time steps.

The problem is a minimization over a set of nested expectations represented in (\ref{eq:multi-stage}).
\begin{subequations} \label{eq:multi-stage}

\begin{equation}
   \begin{aligned} 
    &\min_{\{G_r,F_s,H_s,E_s,e_s^{ini},I_{H2}\}\in\mathcal{C}} \sum_r c_rG_r + \sum_s\left( c_s^f F_s +c_s^hH_s +c_s^eE_s\right) + p_{LTC}I_{H2}
    \\&+ \mathbb{E}_0 \Bigg[ \min_{\substack{\{g_{r,1,h},f_{s,1,h},h_{s,1,h},g^{inf}_{1    ,h},e_{s,1,h}\} \\ \in \mathcal{O}_1(G_r,F_s,H_s,E_s,\xi_1)}} \sum_{h\in \mathcal{H}(1)} \left[\sum_r o_r g_{r,1,h} +vg^{inf}_{1,h}\right]
    + \mathbb{E}_1\bigg[ \cdots \\&+ \mathbb{E}_{T-1} \bigg[ \min_{\substack{\{g_{r,T,h},f_{s,T,h},h_{s,T,h},g^{inf}_{T,h},e_{s,T,h}\} \\ \in \mathcal{O}_T(G_r,F_s,H_s,E_s,e_{s,T-1},\xi_T)}}  \sum_{h\in \mathcal{H}(T)} \left[\sum_r o_r g_{r,T,h} +vg^{inf}_{T,h}\right] + \sum_s c_s^{terminal}\bigg]\bigg] \Bigg]   
\end{aligned} 
\end{equation}

\noindent $c_r$ are the capital costs for generator capacity of type $r$, $c^f_s, c^h_s$ and $c^e_s$ are the capital costs for dispatch power, charging power and energy capacity of storage type $s$ respectively. $p_{LTC}$ is the long-term contract price for hydrogen imports. $G_r$ is the generation capacity choice, $F_s,H_s$ and $E_s$ are the power and energy capacity choices for storage $s$. $I_{H2}$ is the selected long-term contract volume. $\mathcal{C}$ is the compact and convex feasible set for capacity choices given by,
\begin{equation}
    \begin{aligned}
        \mathcal{C}:=\left\{
        \{G_r,F_s,H_s,E_s,e_s^{ini},I_{H2}\} \Bigg| \left\{\substack{0\leq G_r \leq \bar{G_r} \\0\leq F_s\leq \bar{F_s} \\0\leq H_s\leq \bar{H_s}
        \\0\leq E_s\leq \bar{E_s}\\0\leq e^{ini}_s \leq E_s \\0\leq I_{H2}\leq \bar{I}}\right\} \forall r,s
        \right\}.
    \end{aligned}
\end{equation}
\noindent Note that a subset of storage $\mathcal{S}^{LDES}\subset\mathcal{S}$ are long-duration energy storages. The remainder are short-duration storage technologies for which the model does not introduce state variables. In this simplified formalization, we assume that $\mathcal{S}^{LDES}$ is a singleton, consisting of PEM electrolysis, cavern storage and H2 turbines, but an inclusion of e.g. hydro reservoirs would also be possible. The case study above also includes the option of more expensive tank storage.
The dispatch stages $t=1,\dots,T$ are random and enter Eq. \ref{eq:multi-stage} in expectation over the stochastic weather process $\{\xi_t\}_{t=1}^T$. The random weather vector $\xi_t$ affects the feasible space at dispatch stage $t$, $\mathcal{O}_t$. $h\in\mathcal{H}(t)$ is a time step in dispatch stage $t$. $\xi_t$ contains capacity factors for wind, onshore and offshore, and solar PV, $\phi_{r,t,h}$, as well as heat demand $d^{heat}_{t,h}$ and heat pump efficiencies $\psi^{hp}_{t,h}$. $g_{r,t,h}$ is the generation of generator $r$ in dispatch stage $t$ in period $h$. $f_{s,t,h},h_{s,t,h},e_{s,t,h}$ are the corresponding storage dispatch, charging and energy level choices for storage type $s$. $o_r$ are the marginal costs of generator $r$. $g^{inf}_{t,h}$ is a loss of load variable that is priced at the value of lost load (VOLL), $v$. The feasible set of dispatch stage $t$ is given by,
\begin{equation}
    \begin{aligned}
        \mathcal{O}_t(G_r,F_s,H_s,E_s,e_{s,t-1,},\xi_t):=\bigg\{\{g_{r,t,h},f_{s,t,h},h_{s,t,h},g^{inf}_{t,h},e_{s,t,h}\} \bigg| (\ref{eq:energybalance}) - (\ref{eq:penalty})
        \bigg\}
    \end{aligned}
\end{equation}
where,
\begin{align}
    d_{t,h} + (\psi^{hp}_{t,h})^{-1} d_{t,h}^{heat} +  \sum_s h_{s,t,h} - \sum_r g_{r,t,h} - \sum_s f_{s,t,h} = 0 \ &\perp \lambda_{t,h} \in \mathbb{R} \quad \forall h \in \mathcal{H}(t)\label{eq:energybalance} \\
    e_{s,t,h} - e_{s,t,h-1}-\eta_s^h h_{s,t,h} + (\eta_s^f)^{-1}f_{s,t,h} = 0 \ &\perp \lambda_{t,h}^{MSV} \in \mathbb{R} \quad \forall s,h > 1 \label{eq:inner_plant}\\
    -g_{r,t,h}\leq 0   \ &\perp \underline{\mu}_{r,t,h} \geq 0   \quad \forall r,h\\
    g_{r,t,h} - \phi_{r,t,h}G_r \leq 0 \ &\perp \bar{\mu}_{r,t,h} \geq 0  \quad \forall r,h \label{eq:avail} \\
    -f_{s,t,h}\leq 0   \ &\perp \underline{\mu}^f_{s,t,h} \geq 0   \quad \forall s,h\\
    f_{s,t,h} - F_s \leq 0 \ &\perp \bar{\mu}^f_{s,t,h} \geq 0  \quad \forall s,h \\
    -h_{s,t,h}\leq 0   \ &\perp \underline{\mu}^h_{s,t,h} \geq 0   \quad \forall s,h\\
    h_{s,t,h} - H_s \leq 0 \ &\perp \bar{\mu}^h_{s,t,h} \geq 0  \quad \forall s,h \\
    -e_{s,t,h}\leq 0   \ &\perp \underline{\mu}^e_{s,t,h} \geq 0   \quad \forall s,h\\
    e_{s,t,h} - E_s \leq 0 \ &\perp \bar{\mu}^e_{s,t,h} \geq 0  \quad \forall s,h \\
    e_{s,1,1} - e^{ini}_s = 0  &\perp \lambda^{ini}\quad \forall s\in\mathcal{S}^{LDES}
    \\
    e_{s,t,1} -e_{s,t-1,H} -\eta_s^h h_{s,t,1} + (\eta_s^f)^{-1}f_{s,t,1} = 0 \ &\perp \lambda^{plant}_t \in \mathbb{R} \quad \forall t>1,s\in\mathcal{S}^{LDES} \label{eq:plant_eq}
\end{align}

\noindent Note that the weather uncertainty enters each dispatch stage in Eq. \ref{eq:energybalance} for heat demand and in Eq. \ref{eq:avail} for renewable capacity factors. We assume that $\phi_{r,t,h}$ are constant for $r\in\{bio,nuclear\}.$ 

The stages are connected by the capacity choices in stage $t=0$, $G_r,H_s,F_s$ and $E_s$ and the initial storage level $e_s^{ini}$ as well as the plant equation Eq. \ref{eq:plant_eq} propagating the storage level through the stages. The plant equation only needs to hold for storages $s\in\mathcal{S}^{LDES}$. The remaining storage technologies are subject to a circularity constraint within each dispatch stage that is omitted here for simplicity. The Greek symbols following the perpendicular sign, $\perp$, are the dual variables to the constraints. We use them in the derivation of the storage bidding curve in Section \ref{sec:bidding_theory}.  

In the final stage $T$ the stage objective contains a terminal cost term, which we define with the help of an auxiliary variable $p_T \in \mathbb{R}_+$,
\begin{align}
       e_s^{ini} - e_{s,T,H} - p_{s,T} \leq 0 \ &\perp \rho^{pen} \geq 0 \quad \forall s\in\mathcal{S}^{LDES} \label{eq:penalty}\\
    c^{terminal}_s &:= v p_{s,T} \label{eq:terminal_cost}
\end{align}

An optimal policy determines an initial storage level for each storage type $s\in\mathcal{S}^{LDES}$. Any negative deviation at the end of the time horizon is penalized at VOLL.
\end{subequations}

\subsubsection{The Perfect Foresight Model}

The Perfect Foresight Model which serves as a benchmark in isolating the effect of limited foresight under weather uncertainty, is a two-stage stochastic program and a special case of the multi-stage stochastic program discussed in the previous subsection. The problem stages are $t=0,1$, where stage $0$ is the capacity stage and stage $1$ is the dispatch stage. Stage $1$ has 2190 four-hourly time steps to cover an entire weather year. The number of scenarios is $N=N_1$, equal to 35 in the case study. Capacity decisions and the choice of the initial storage level are taken under uncertainty of which weather year is going to materialize in the dispatch stage. Otherwise, the system is operated under perfect foresight. 

\subsection{Primer to SDDP}\label{sec:sddp}

This section provides a very brief introduction to the stochastic dual dynamic programming (SDDP) algorithm used to solve the Problem in Eq. \ref{eq:multi-stage} for the Limited Foresight model in the main body of the text. The section builds on Dowson (2020) \cite{dowson_policy_2020} and Shapiro (2011) \cite{shapiro_analysis_2011} as well as Shapiro et al. (2021) \cite{shapiro_lectures_2021}, and we refer to these texts for a more thorough treatment.

We begin by re-introducing the dynamic programming equation from Section \ref{sec:theory}.
\begin{equation}\label{eq:si_Bellman}
 \begin{aligned}
     Q_t(e_{s,t-1,H},G_r,F_s,H_s,E_s,e^{ini}_s,\xi_{t}):=& \\\min_{\substack{g_{r,t,h},f_{s,t,h},h_{s,t,h},g^{inf}_{t,h},e_{s,t,h} \\ \in \mathcal{O}_t(G_r,F_s,H_s,E_s,e_s^{ini},e_{s,t-1},\xi_{t})}}  \sum_h\left[\sum_r o_r g_{r,t,h} +vg^{inf}_{t,h}\right] &+ \mathbb{E}_t\left[{Q}_{t+1}(e_{s,t,H},G_r,F_s,H_s,E_s,e_s^{ini})\right]
\end{aligned}
\end{equation}

We make several necessary assumptions:
\begin{enumerate}
    \item \textit{Relatively complete recourse}: We assume that regardless of decisions in stages $1,\dots,t-1$ and of the realization of $\xi_t$, there exists a feasible solution in stage $t$, i.e. $\mathcal{O}_t\neq\emptyset$
    \item \textit{Stagewise independence}: As in the main text, we assume that the stochastic process is \textit{stagewise independent}. That is, the distribution of $\xi_t$ does not depend on $\xi_s$ for $s=1,\dots,t-1$.
    \item \textit{Finite state space}: We assume that the data process $\Xi$ is finite. One can think of the solution to our problem as a sample average approximation of the true problem over the true continuous weather variables.
\end{enumerate}

Assuming that each stage has a number of samples $N_t$, the total number of scenarios given the stagewise independence assumption is $N=\prod_{t=1}^T N_t$. Based on assumption 3, we replace the expectation operator in \ref{eq:si_Bellman} by $\mathcal{Q}_t(\cdot)=\frac{1}{N_t}\sum_i^{N_t}Q_t(\cdot,\xi_{t,i}), t = 1,\dots,T$.

By linearity of the subproblems, the cost-to-go functions $\mathcal{Q}_t(\cdot)$ are convex, piecewise-linear, in the state variables. We decompose the model by replacing the expected cost-to-go in each stage $t$ by a lower approximation of the same, a cutting-plane model. For simplicity lets subsume all outgoing state variables in a vector-valued variable $x_t=\{e_{s,t,H},G_r,F_s,H_s,E_s,e_s^{ini}\}$ and all dispatch variables as $y_t =\{g_{r,t,h},f_{s,t,h},h_{s,t,h},g^{inf}_{t,h},e_{s,t,h}\}$. Given an arbitrary iteration $k$ of the algorithm, a weather realization $\xi_{t,i}$ and an incoming state $x_{t-1}$, we can write the subproblem in stage $t$ as,
\begin{subequations} \label{eq:approx_sp}
    
\begin{align}
   \hat{Q}^k_t(x_{t-1},\xi_{t,i}) = \min_{y_t,x_t} c_t^Ty_t + \theta_{t}
\end{align}
s.t. 
\begin{align}
    y_t &\in \mathcal{O}_t(\bar{x},\xi_t) \label{eq:sub_feasible}\\ 
    \bar{x} &= x_{t-1} \perp \rho_t \label{eq:fishing_constraint}\\
    x_t &= T(y_t,x_{t-1})\label{eq:state_transition}\\
    \theta_t &\geq \frac{1}{N_{t+1}}\sum_{i=1}^{N_{t+1}}\left[\hat{Q}^j_{t+1} (x^j_t,\xi_{t+1,i}) + (\rho^j_{t+1})^T(x_t-x_t^j ) \right], j = 1,\dots, k \label{eq:cpm}\\
    \theta_t &\geq 0 \label{eq:lower_bound_ctg}
\end{align}

\noindent Note that we have replaced the cost-to-go term by $\theta_t$, an approximation. We also simplify the representation of the current stage's dispatch costs by $c_t^Ty_t$. The cost vector $c_t$ only has non-zero entries for dispatchable generation with marginal generation costs and for load-shedding (or imports in the case study). Eq. \ref{eq:sub_feasible} defines the feasible space for dispatch decisions $y_t$ given the weather realization $\xi_{t,i}$ and incoming state variable vector $x_{t-1}$. We define a \textit{fishing} constraint Eq. \ref{eq:fishing_constraint}  to conveniently collect the vector of duals with respect to the incoming state variables, which we define as $\rho_t$ \cite{dowson_policy_2020}. Eq. \ref{eq:state_transition} describes the transition of state variables. For capacities and the initial storage level $e_s^{ini}$ functional $T(\cdot)$ just passes on without altering anything. The transition from $e_{s,t-1,H}$ to $e_{s,t,H}$ is defined by Eq. \ref{eq:plant_eq} and \ref{eq:inner_plant}. Eq. \ref{eq:cpm} describes the cutting plane model. $\theta_t$ is bounded from below by a collection of cutting planes or \textit{cuts}, defined by subgradients of $\mathcal{Q}_{t+1}$ obtained in previous iterations $j=1,\dots,k$.\footnote{Since the cost-to-go functions are piecewise-linear they are not differentiable and we require the generalization of a gradient, that is a subgradient. See \cite{goke_stabilized_2024} for details.} The last constraint just ensures that our cost estimate is non-negative.

The more cuts we add to the approximate subproblem, the closer $\theta_t$ resembles $\mathcal{Q}_{t+1}(\cdot)$ and the approximation converges to the true cost-to-go function. Next, we discuss how to add new cuts in an iteration. Each iteration consists of two steps, a forward pass and a backward pass.

\end{subequations}

\subsubsection{The forward pass}

We need to create trial points $\{x_t^k\}_{t=0}^T$. At the start of iteration $k$, we create a random sample path $\xi_i$ and begin by solving the approximate subproblem at stage $0$. We pass on $x^k_0$ to stage $1$, observe realization $\xi_{1,i}$ and solve the approximate subproblem in Eq. \ref{eq:approx_sp}, pass on $x^k_1$ to stage $2$, observe $\xi_{2,i}$ and so on. Note that the \textit{relatively complete recourse} assumption ensures that the next stage's subproblem remains feasible regardless of the incoming state variable or the weather realization.

\subsubsection{The backward pass}

Based on the forward pass, we obtain a vector of outgoing states $\{x_t^k \}$. In the backward pass, we start in stage $T-1$ since the stage $T$ does not have a cost-to-go term. In stage $T-1$, we take candidate state ${x_{T-1}^k}$ and solve the subproblem in stage $T$, ${Q}_T(x^k_{T-1},\xi_{T,i})$ for all $\xi_{T,i}\in\Xi_T$. For each weather realization $i$, we compute the dual value $\rho^k_T(x_{T-1}^k,\xi_{T,i})$ and evaluate ${Q}_T(x^k_{T-1},\xi_{T,i})$ to construct a new cut in $T-1$ according to Eq. \ref{eq:cpm}.

Next we move to stage $T-2$, take the candidate outgoing state $x_{T-2}^k$ and solve the \textit{approximate} subproblem for $T-1$,  $\hat{Q}_{T-1}(x^k_{T-2},\xi_{T-1,i})$ for all $\xi_{T-1,i}\in\Xi_{T-1}$. Again, we compute dual variables $\rho_{T-1}^k(x_{T-2}^k,\xi_{T-1,i})$ and cost-to-go $\hat{Q}_{T-1}(x^k_{T-2},\xi_{T-1,i})$ for all weather realizations and a add a new cut to the subproblem in stage $T-2$. Note that in this stage, the next stage's subproblem is the approximate model $\hat{Q}^k_{T-1}$ that has already been updated with cuts in the previous step of the backward pass. We continue the procedure until stage $0$.

\subsubsection{Policy, Bounds and termination}

Given a model trained for $K$ iterations, we can define a \textit{policy} $z^K(\xi) = \{x^K_0,x^k_t(\xi_t),y^K_t(\xi_t)\}_{t=1}^T$, $x_t(\xi_t)$ is the outgoing state vector and $y_t(\xi_t)$ is the optimal dispatch after solving $\hat{Q}^K(x^K_{t-1},\xi_t)$.

A lower bound to the true optimal policy value is given by the (approximate) investment stage objective,
\begin{align}
LB=min_{x_0\in\mathcal{C}} c_0^T x_0 + \hat{\mathcal{Q}}^K_1(x_0)
\end{align}

Stochastic dual dynamic programming does not have a deterministic upper bound as we cannot evaluate the entire state space (recall that the number of scenarios is a very large number $N$). We can simulate a Monte Carlo estimator of the upper bound instead. To this end, we simulate a collection of sample paths ${\xi_i}, \ i\in \mathcal{M}$, where $|\mathcal{M}|<<N$ and simulate forward passes on these paths to calculate a Monte Carlo estimator for the upper bound,

\begin{align}
   \hat{UB} =  c_0^Tx^K_0 +\frac{1}{|\mathcal{M}|}\sum_{i\in\mathcal{M}}\left[ \sum_{t=1}^T c_t^Ty^K_t(\xi_{t,i})\right]
\end{align}

The classical termination rule for SDDP would be to stop once the lower bound is within the confidence interval of the upper bound \cite{pereira_multi-stage_1991}. However, this rule has its weaknesses \cite{dowson_policy_2020}, and termination is an open research topic.

In the case study above, we opt to set a fixed iteration limit of $K=15000$ and inspect the convergence of capacity variables and a stalling of the lower bound ex-post in the next section.

\subsubsection{Simulation of historical weather years}

To facilitate a comparison of the dispatch behavior between the Limited and the Perfect Foresight models, we use the resulting policy to simulate Limited Foresight in the historical weather years in the sample. First, note that the set of historical weather years $\Xi^{hist}$ is a comparably small subset of $\Xi$ with $N^{hist}=35$ in the case study above. For each historical weather year $i\in \Xi^{hist}$, we compute the solution to a forward pass. The capacities remain the same for all historical weather years, and the dispatch decisions result from the trained policy $\{y^K_t(\xi_{t,i})\}_{t=1}^T$.

\subsubsection{Computing marginal storage values}

Lastly, we discuss how to derive the marginal storage values or long-duration energy storage bidding curves from the trained model in Section \ref{sec:results_ldes_bidding}.

Given our trained policy, we have for each stage $t=1,\dots,T-1$ optimal value function approximations $\hat{\mathcal{Q}}^K_t(x_{t-1})$. Based on Eq. \ref{eq:approx_sp}, the marginal value of adding a unit of an element of $x_{t-1}$ is equal to the corresponding element of $\rho_t$, i.e.,
\begin{align*}
    \frac{\partial \mathcal{Q}_t^K(x_{t-1})}{\partial x_{t-1}} = \rho'(x_{t-1}) =  N_t^{-1}\sum_{i=1}^{N_t} \rho'_t(x_{t-1},\xi_{t,i}) 
\end{align*}
\noindent where we define $\rho'$ to be the slope the maximum cut at $x_{t-1}$ after $K$ iterations. Let $\lambda^{MSV}_{t,1}$ is the element of $\rho'(x_{t-1})$ corresponding to $e_{s,t-1,H}$ in the state vector $x_{t-1}$.

\subsection{Model convergence} \label{sec:convergence_details}

\begin{figure}[H]
    \centering
    \includegraphics[width=0.8\linewidth]{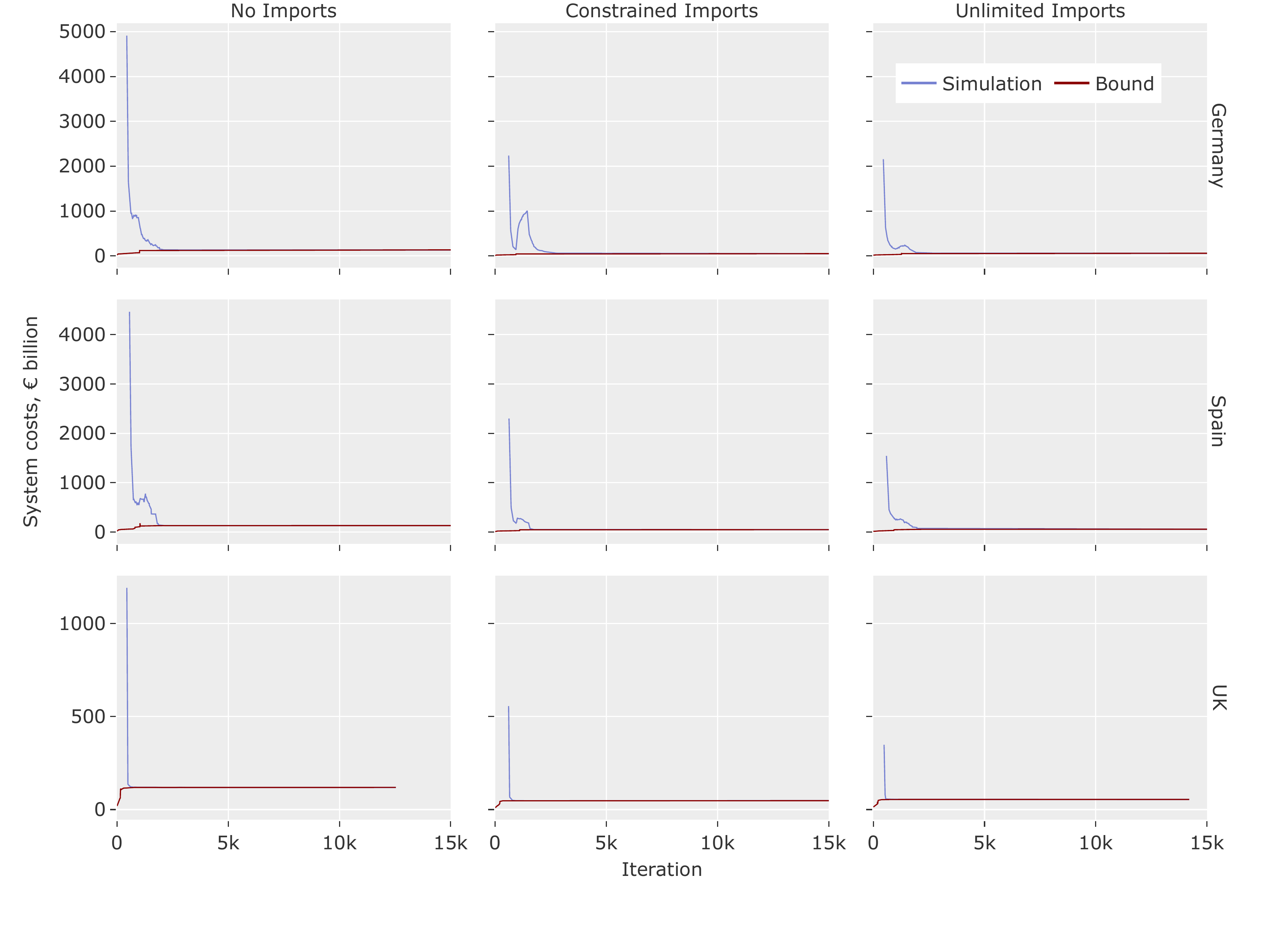}
    \caption{\textit{Objective value convergence}: Convergence of the lower bound corresponding to the model's first stage objective value and the 100 iterations rolling mean simulations as an estimator for the upper bound for all 9 model specifications.}
    \label{fig:convergence}
\end{figure}

This section provides some details on model convergence for the Limited Foresight model solved with stochastic dual dynamic programming (SDDP). Figure \ref{fig:convergence} shows the convergence of the model's lower bound and a rolling average of forward pass simulations for all model specifications. 

\begin{figure}[H]
    \centering
    \includegraphics[width=0.9\linewidth]{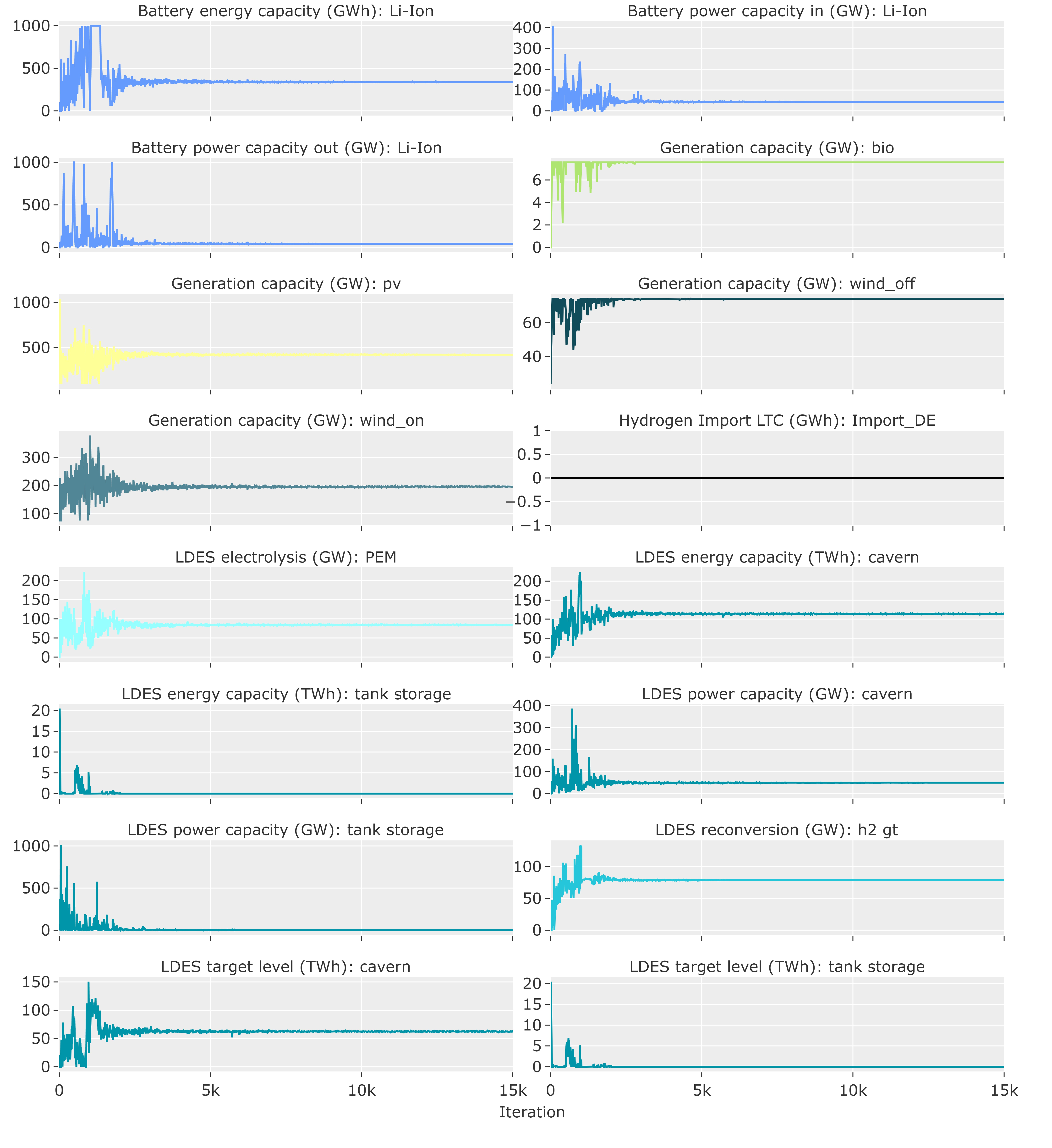}
    \caption{\textit{Capacity convergence}: Evolution of capacity choices by SDDP iteration (example for Germany under no H2 spot imports). All technology choices stabilise early on. Some variations persist until late in the training process. Continued training ensures that the model has seen more system states resulting in a better storage policy even if capacities no longer change.}
    \label{fig:capa_evo}
\end{figure}

Table \ref{tab:computational_res} provide run times and iteration number for all specifications and Figure \ref{fig:capa_evo} shows the convergence of the capacity choices in stage 0.

\begin{table}[]
    \centering
    \resizebox{\textwidth}{!}{\begin{tabular}{rrrrrr}
  \hline
  \textbf{Country} & \textbf{Scenario} & \textbf{Iterations} & \textbf{Time (h)} & \textbf{Total solves (millions)} & \textbf{Active stopping rule} \\\hline
  Germany & No H2 spot imports & 15003 & 50.69 & 6.5113 & Iteration limit \\
  Spain & No H2 spot imports & 15003 & 38.8 & 6.5113 & Iteration limit \\
  United Kingdom & No H2 spot imports & 15003 & 51.59 & 6.5113 & Iteration limit \\
  Germany & Constr. H2 spot imports & 15003 & 35.81 & 6.5113 & Iteration limit \\
  Spain & Constr. H2 spot imports & 15003 & 41.92 & 6.5113 & Iteration limit \\
  United Kingdom & Constr. H2 spot imports & 15003 & 39.1 & 6.5113 & Iteration limit \\
  Germany & Unlimited H2 spot imports & 12537 & 54.02 & 5.4411 & Time limit \\
  Spain & Unlimited H2 spot imports & 15003 & 43.59 & 6.5113 & Iteration limit \\
  United Kingdom & Unlimited H2 spot imports & 14203 & 54.02 & 6.1641 & Time limit \\\hline
\end{tabular}
}
    \caption{\textit{Computational details of the SDDP implementation by Scenario}}
    \label{tab:computational_res}
\end{table}

\subsection{Additional model outputs}
\begin{figure}[H]
    \centering
    \includegraphics[width=\linewidth]{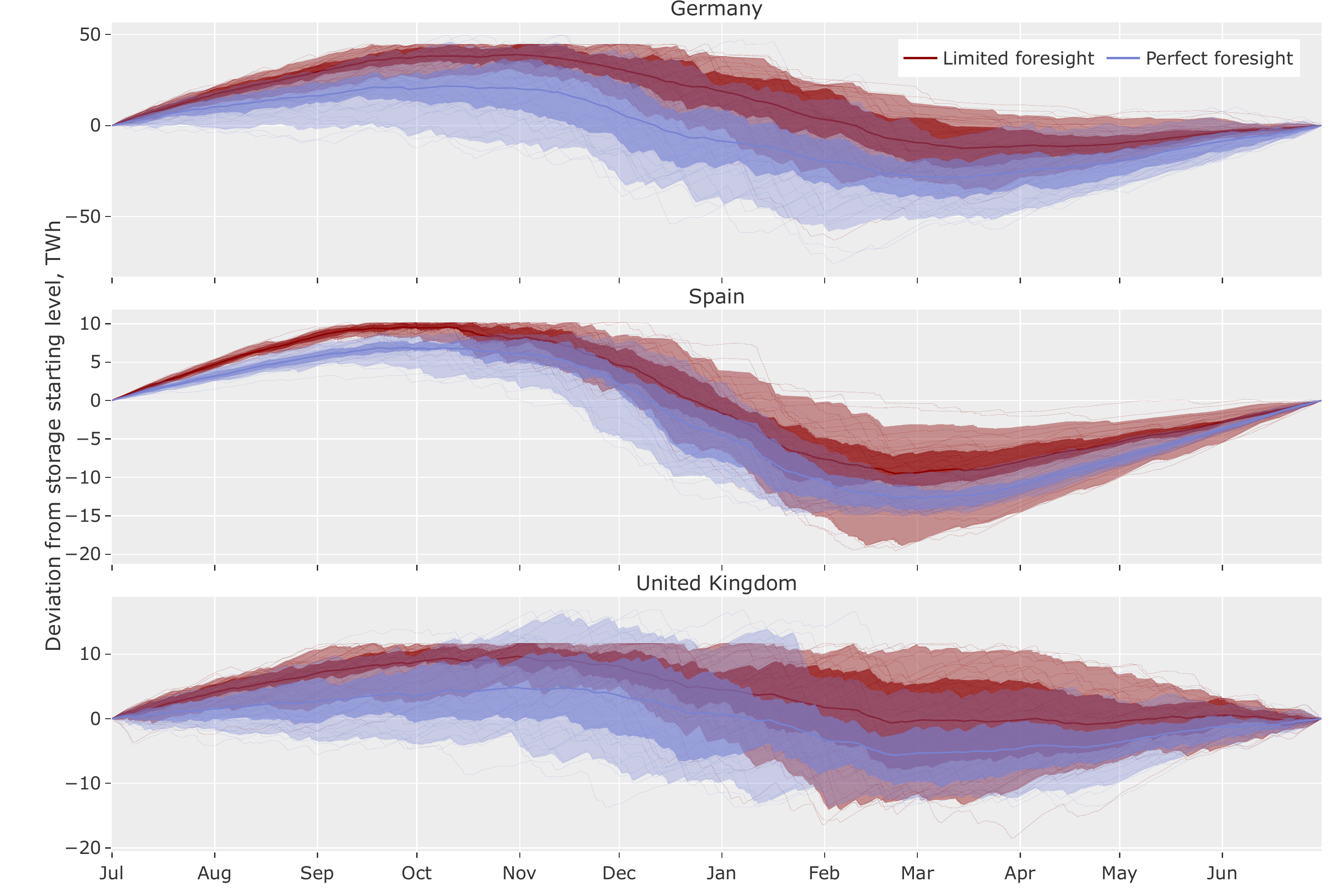}
    \caption{\textit{LDES trajectories in Constrained Imports}}
    \label{fig:traj_constr}
\end{figure}

\begin{figure}[H]
    \centering
    \includegraphics[width=\linewidth]{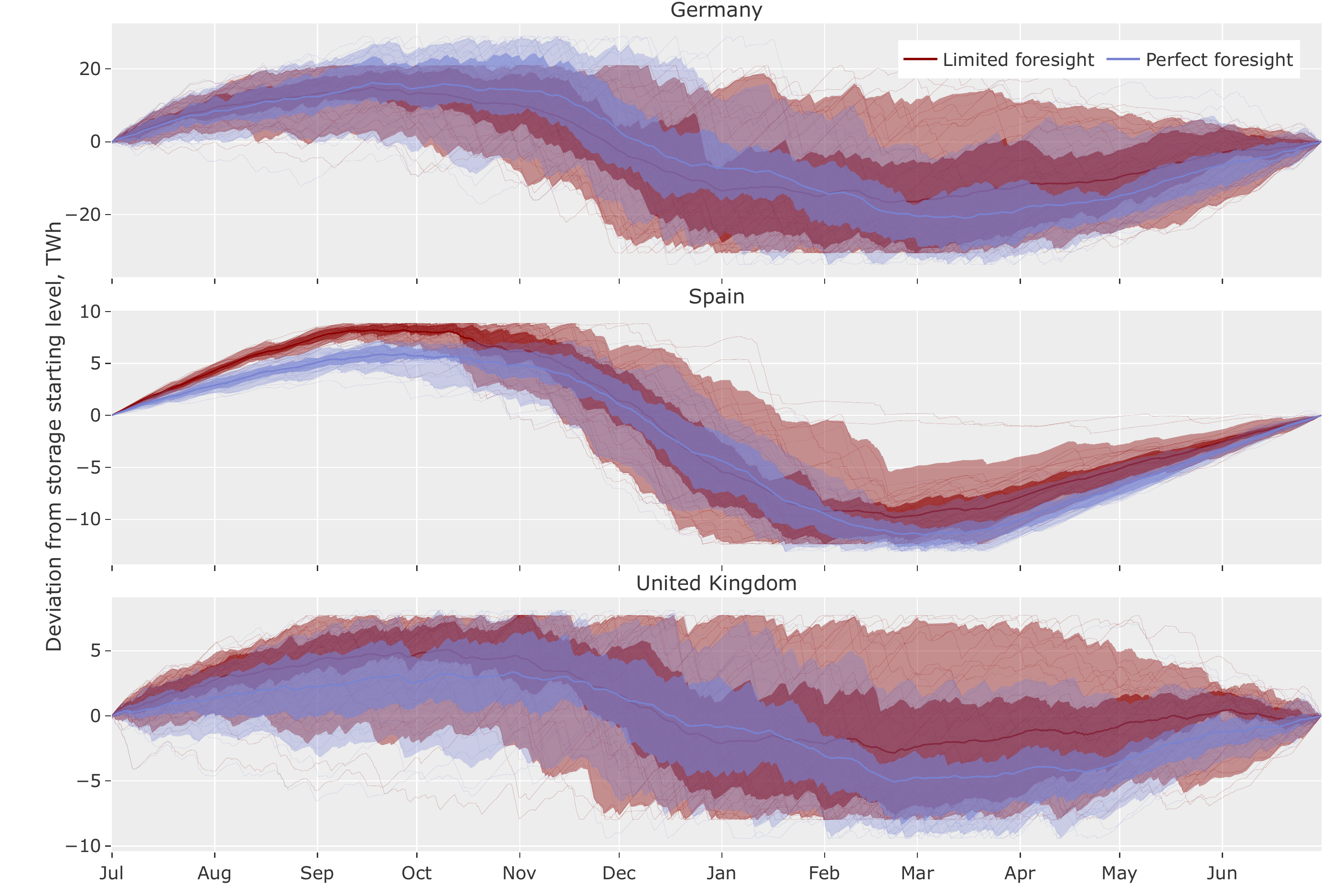}
    \caption{\textit{LDES trajectories in Unlimited Imports}}
    \label{fig:traj_unlimit}
\end{figure}

This section presents additional results. Specifically, Figures \ref{fig:traj_constr} and \ref{fig:traj_unlimit} replicate Figure \ref{fig:ldes_traj}.

\begin{figure}[H]
    \centering
    \includegraphics[width=\linewidth]{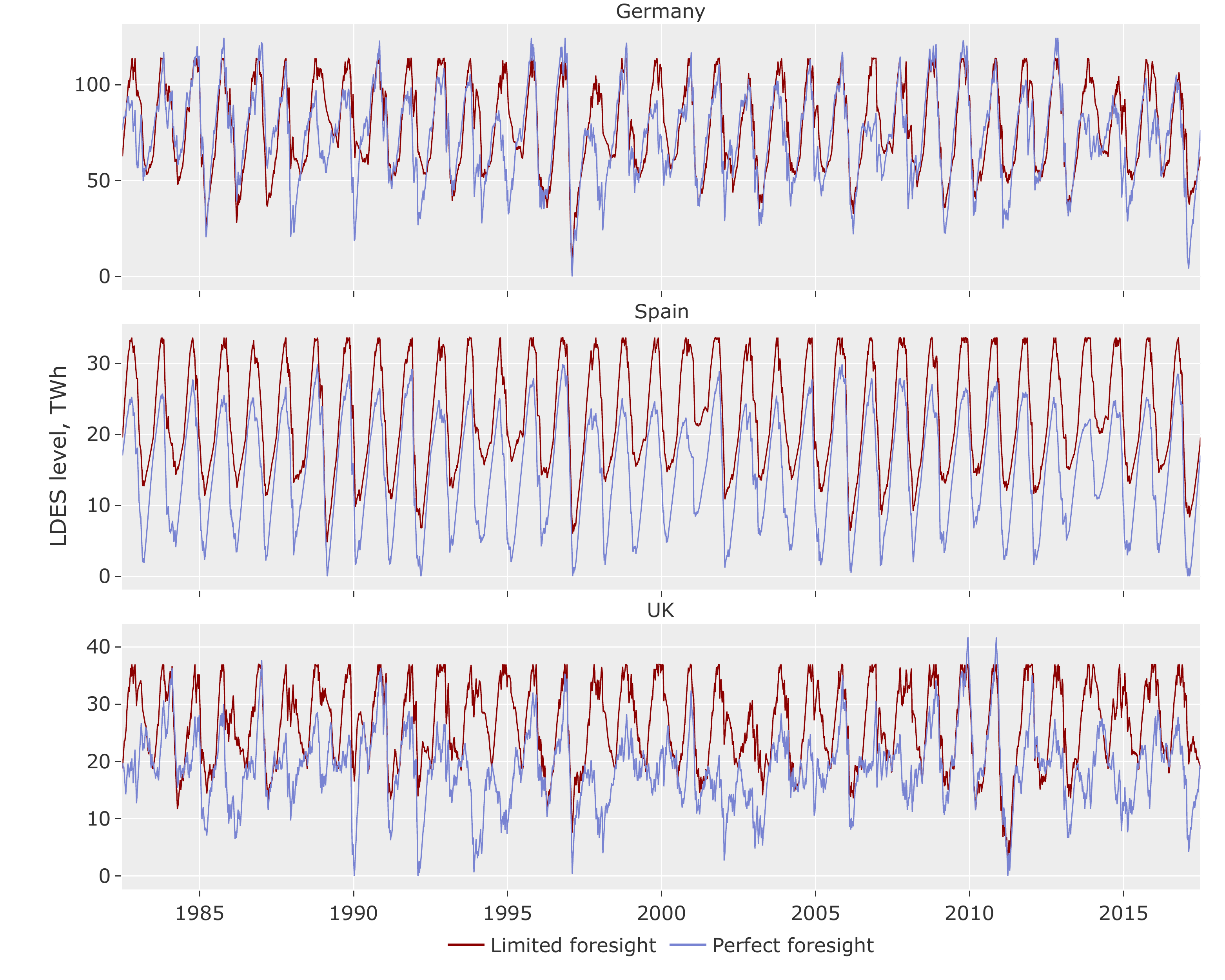}
    \caption{\textit{LDES trajectories in chronological order in No imports}}
    \label{fig:chrono_traj}
\end{figure}

Figure \ref{fig:chrono_traj} shows the storage trajectories for 35 historical weather years in chronological sequence.

\begin{figure}[H]
    \centering
    \includegraphics[width=0.9\linewidth]{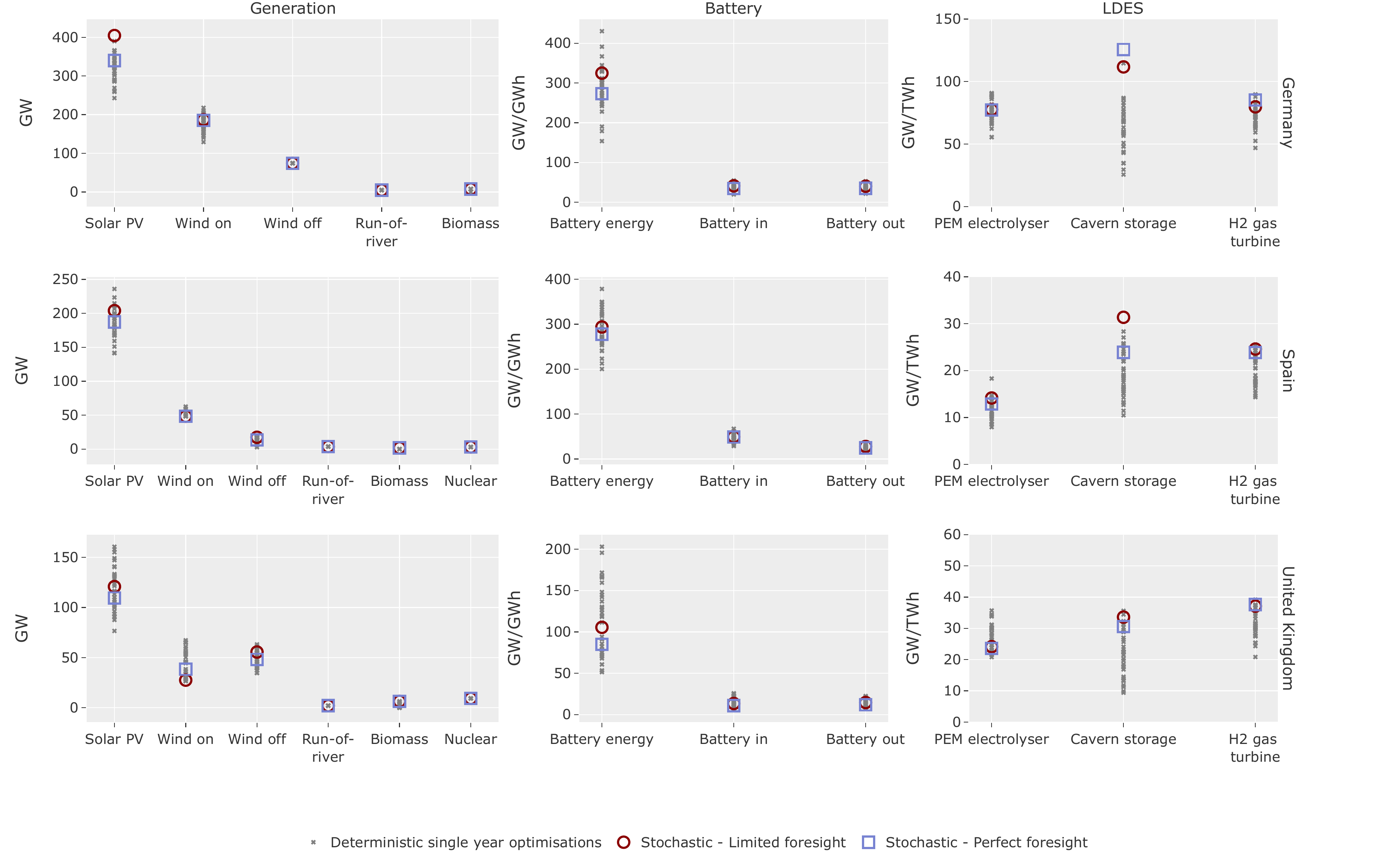}
    \caption{\textit{Optimal capacities  - Constrained Imports}}
    \label{fig:capa_constr_outside}
\end{figure}

\begin{figure}[H]
    \centering
    \includegraphics[width=0.9\linewidth]{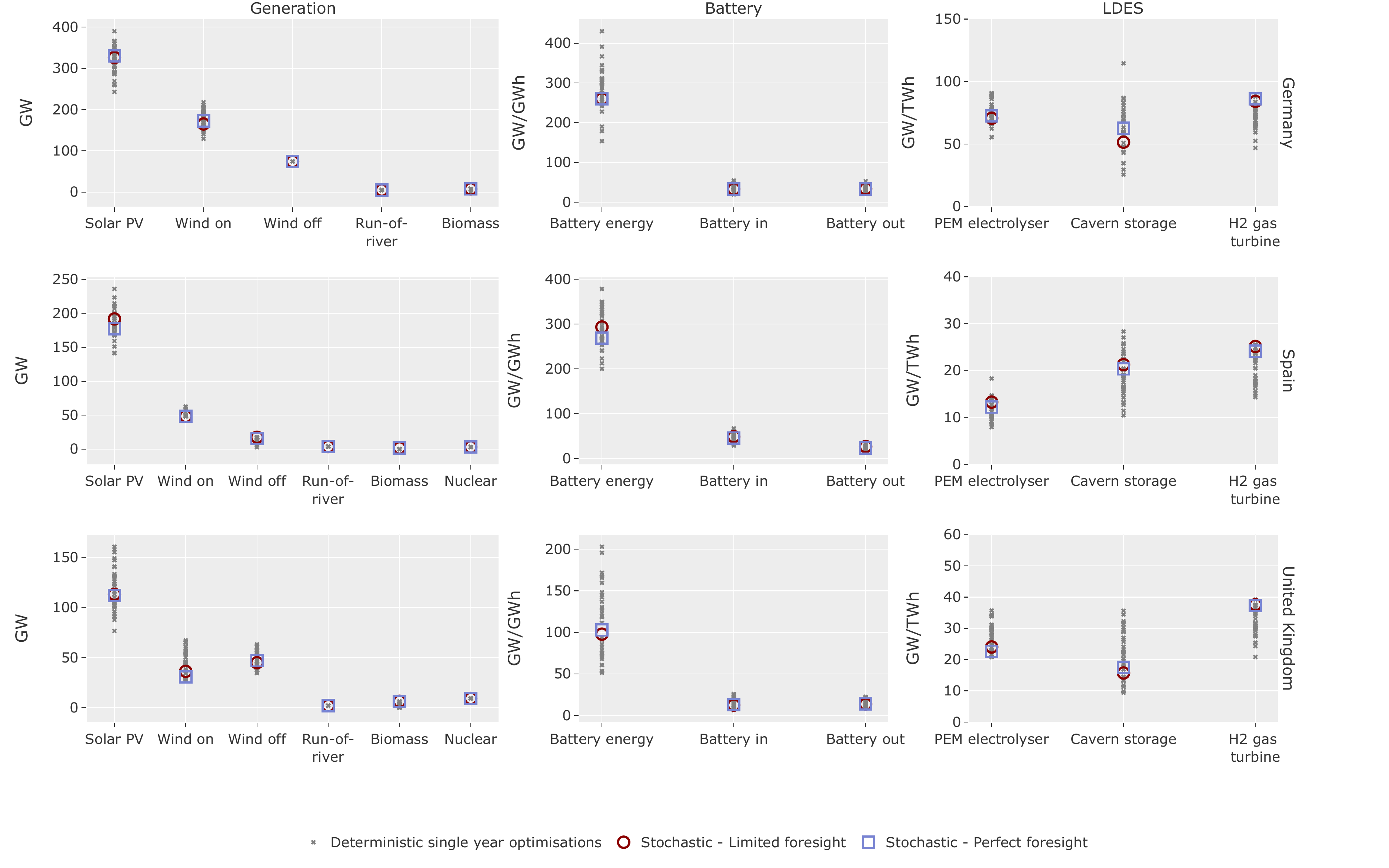}
    \caption{\textit{Optimal capacities - Unlimited Imports}}
    \label{fig:capa_outside}
\end{figure}

Figures \ref{fig:capa_constr_outside} and \ref{fig:capa_outside} show the capacity decisions under Perfect and Limited Foresight for the \textit{Constrained Imports} and \textit{Unlimited Imports} scenarios.

\begin{figure}[H]
    \centering
    \includegraphics[width=0.9\linewidth]{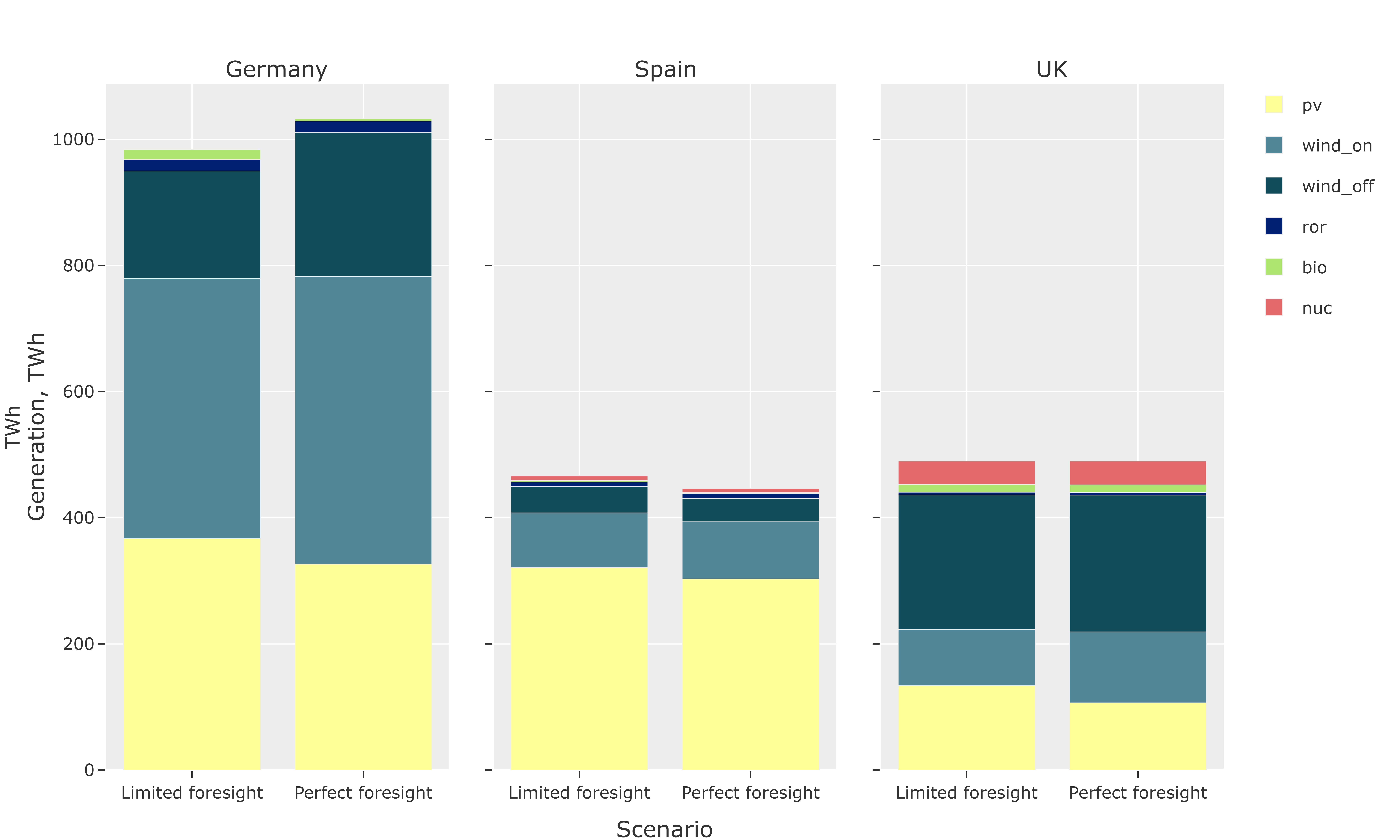}
    \caption{\textit{Differences in generation mixes between PF and LF (No Imports)}}
    \label{fig:gen_diff}
\end{figure}

\begin{figure}[H]
    \centering
    \includegraphics[width=0.9\linewidth]{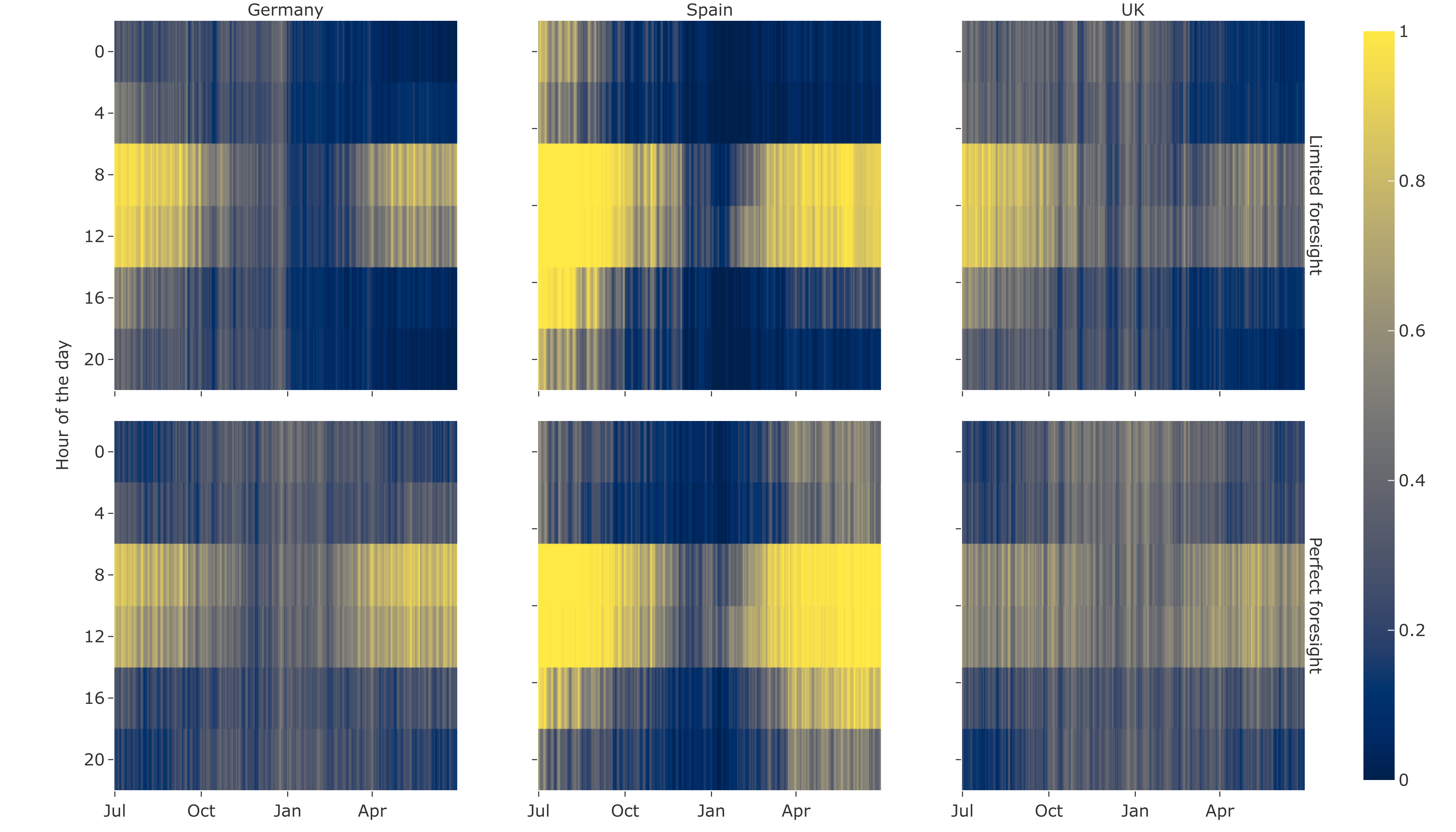}
    \caption{\textit{Mean hourly electrolysis capacity factor by time of day (No Imports)}}
    \label{fig:hourly_hm}
\end{figure}

\begin{table}[H]
    \centering
    \begin{tabular}{rrrr}
  \hline
   \textit{Limited foresight} &&&\\& \textbf{Solar PV} & \textbf{PEM electrolysis} & \textbf{Onshore wind}\\\hline
\textbf{Solar PV} & 1.0 & 0.50 & -0.37 \\
 \textbf{PEM electrolysis} & 0.50 & 1.0 & 0.31 \\
 \textbf{Onshore wind} & -0.37 & 0.31 & 1.0 \\\hline
   \textit{Perfect foresight} &&&\\& \textbf{Solar PV} & \textbf{PEM electrolysis} & \textbf{Onshore wind}\\\hline
\textbf{Solar PV} & 1.0 & 0.43 & -0.28 \\
 \textbf{PEM electrolysis} & 0.43 & 1.0 & 0.55 \\
 \textbf{Onshore wind} & -0.28 & 0.55 & 1.0 \\\hline
\end{tabular}
    \caption{\textit{Correlation of pv and wind generation with PEM electrolysis by scenario (No Imports)}}
    \label{tab:corr_mat}
\end{table}

\begin{figure}[H]
    \centering
    \includegraphics[width=0.9\linewidth]{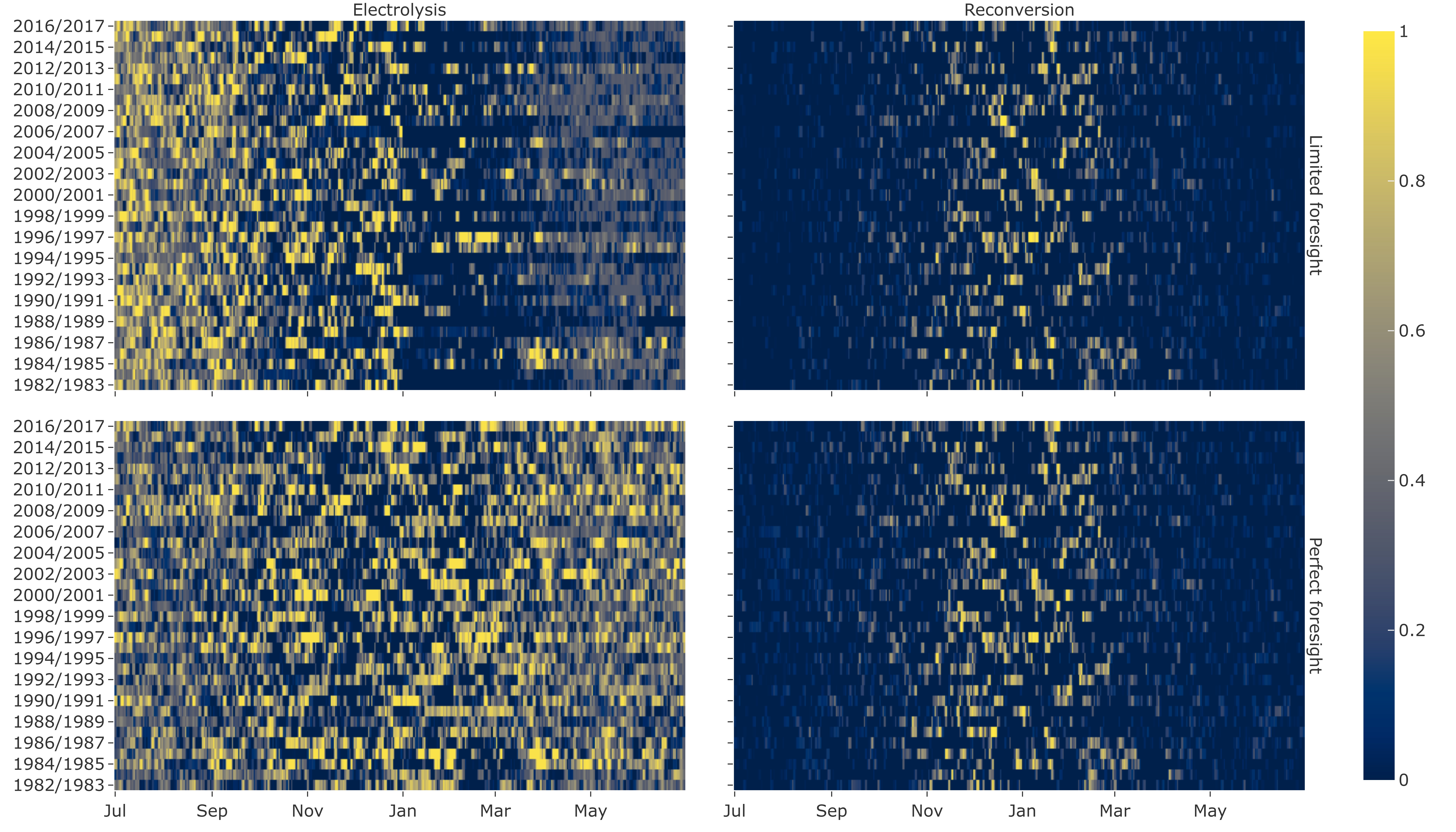}
    \caption{\textit{Daily electrolysis and reconversion capacity factors across all weather years (No Imports)}}
    \label{fig:ely_recon_hm}
\end{figure}
\begin{figure}[H]
    \centering
    \includegraphics[width=0.9\linewidth]{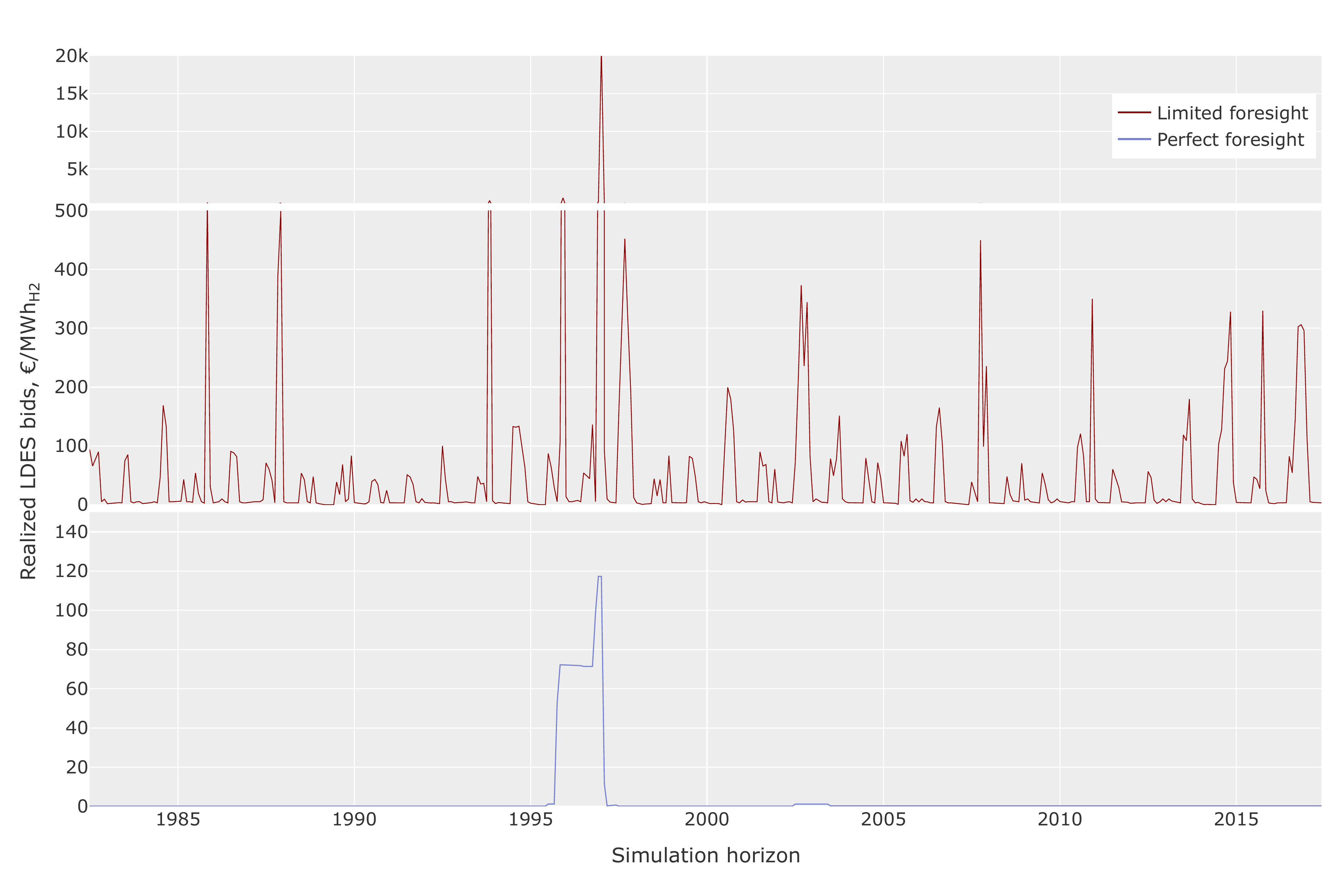}
    \caption{\textit{Realized marginal storage values for Germany over the simulation horizon under Perfect Foresight and Limited Foresight - No Imports}}
    \label{fig:msv_realized_no_imports}
\end{figure}
\begin{figure}[H]
    \centering
    \includegraphics[width=0.9\linewidth]{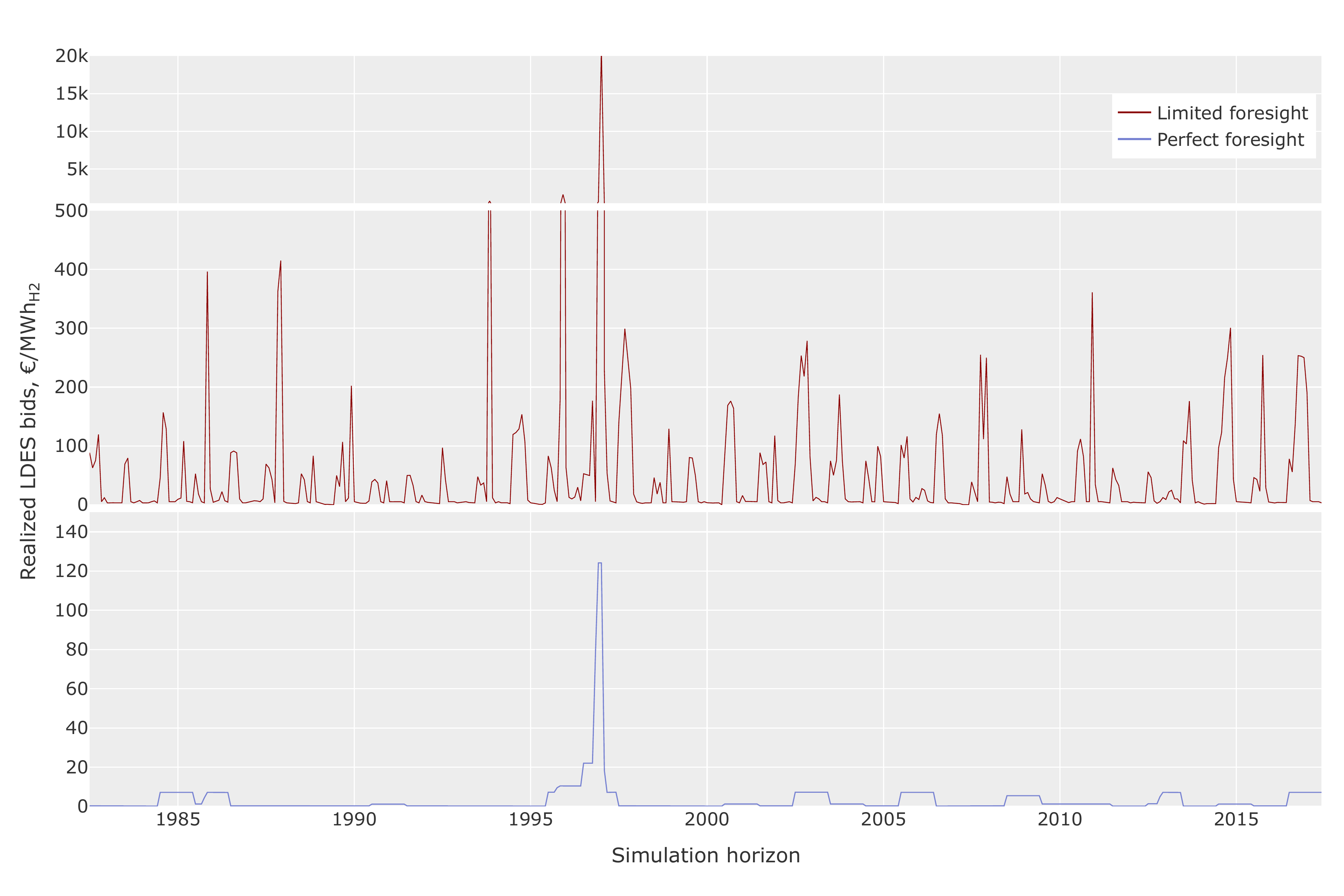}
    \caption{\textit{Realized marginal storage values for Germany over the simulation horizon under Perfect Foresight and Limited Foresight - Constrained Imports}}
    \label{fig:msv_realized_constr_imports}
\end{figure}
\begin{figure}[H]
    \centering
    \includegraphics[width=0.9\linewidth]{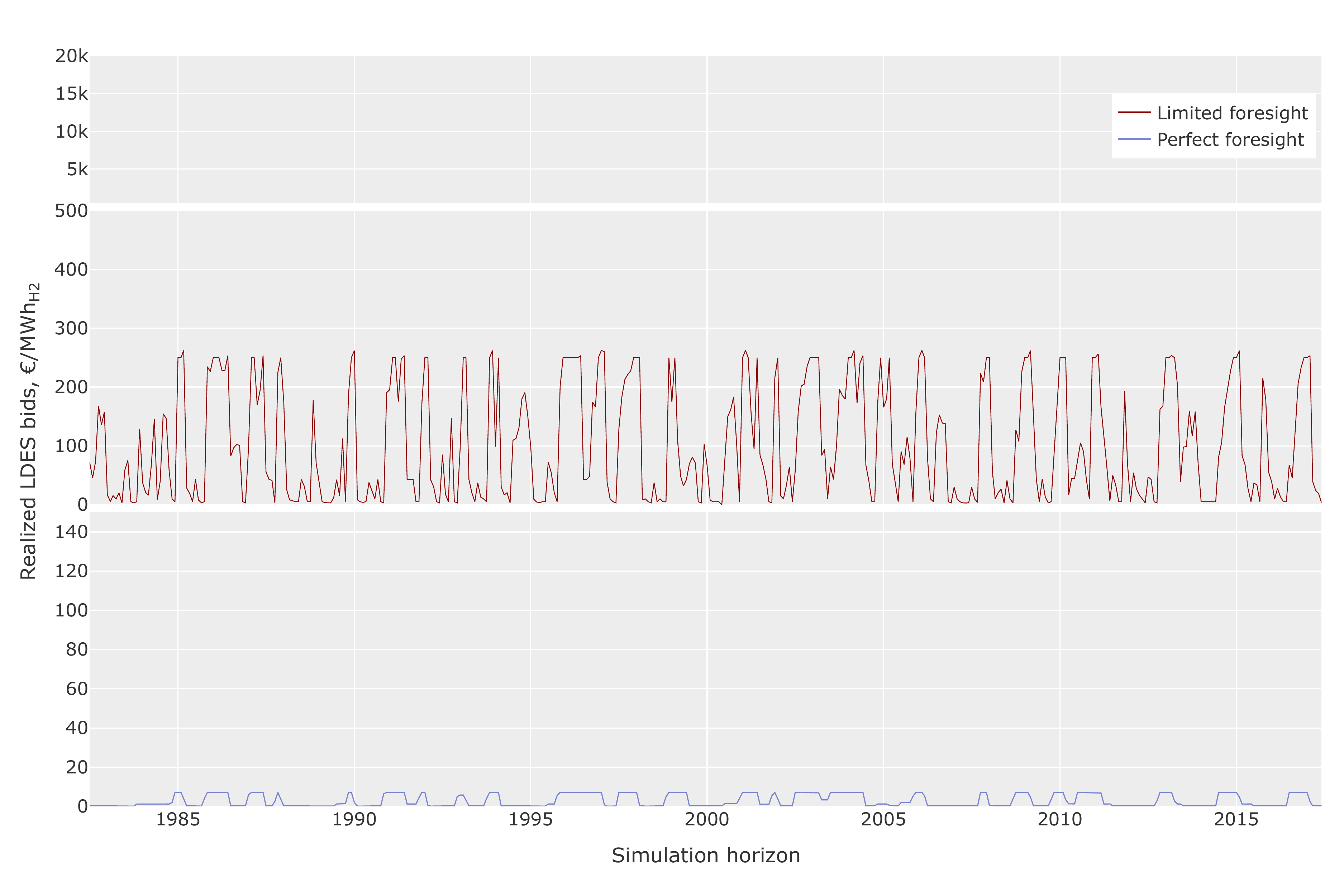}
    \caption{\textit{Realized marginal storage values for Germany over the simulation horizon under Perfect Foresight and Limited Foresight - Unlimited Imports}}
    \label{fig:msv_realized_unlimited}
\end{figure}

Figures \ref{fig:msv_realized_no_imports} to \ref{fig:msv_realized_unlimited} show the realized marginal storage values for the simulation period.

\subsection{Input data}
\begin{figure}[H]
    \centering
    \includegraphics[width=0.7\linewidth]{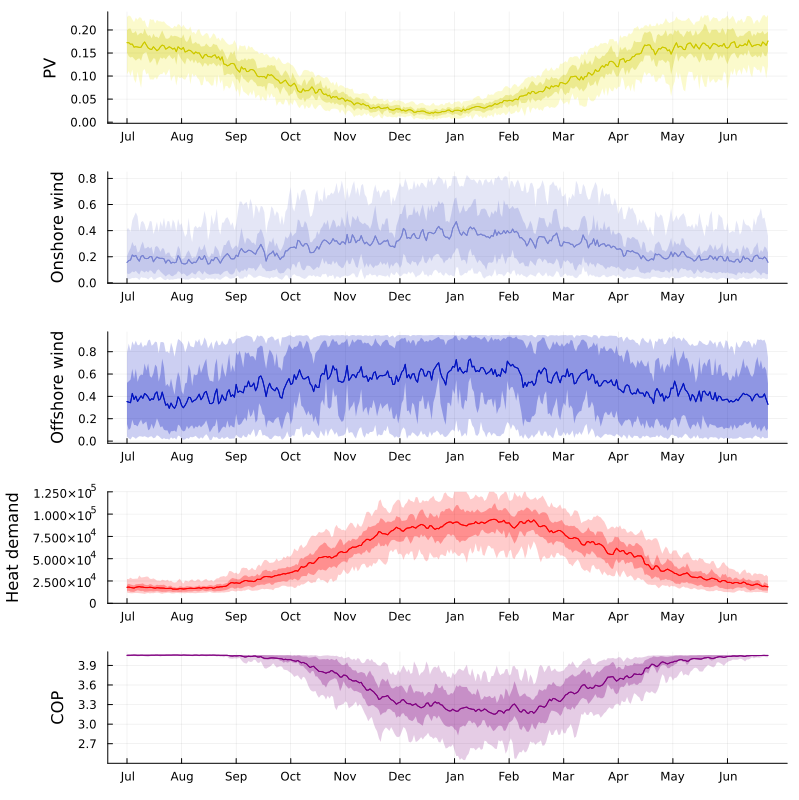}
    \caption{\textit{Weather data distribution}}
    \label{fig:input_dist}
\end{figure}

\begin{table}[H]
    \centering
    \resizebox{\textwidth}{!}{\begin{tabular}{rrrrrrrrrr}
  \hline
  \textbf{Technology} & \textbf{Investment costs} & & \textbf{Fixed O\&M} &  & \textbf{Variable + Fuel} &  & \textbf{Efficiency} & \textbf{Lifetime} \\ \hline
  Biomass & 2516.96 & €/kW & 129.01 & €/kW/a & 13.6 & €/MWh & 49\% & 30  \\
  Solar PV & 457.84 & €/kW & 8.76 & €/kW/a &  &  & - & 40  \\
  Onshore wind & 1224.98 & €/kW & 17.53 & €/kW/a & 2.1 & €/MWh & - & 30  \\
  Offshore wind & 1842.61 & €/kW & 37.08 & €/kW/a & 3.4 & €/MWh & - & 30 \\
  Battery inverter & 71.7 & €/kW & 0.65 & €/kW/a & 2.0 & €/MWh & 96\% & 30  \\
  Battery storage & 112.31 & €/kWh &  &  &  &  & 100\% & 30  \\
  PEM electrolysis & 404.22 & €/kW & 8.05 & €/kW/a &  &  & 66\% & 25/20 &  \\
  H2 turbine & 501.38 & €/kW & 7.89 & €/kW/a & 5.0 &€/MWh  & 43\% & 25 \\
  H2 cavern & 1.43 & €/kWh &  &  &  &  & 100\% & 100  \\
  H2 cavern compressor & 95.66 & €/kW & 3.83 & €/kW/a &  &  & 100\% & 15 &  \\
  H2 tank storage & 17.15 & €/kWh & 0.48 &  &  &  & 100\% & 30  \\
  H2 tank storage compressor & 9.55 & €/kW & 0.48 & €/kW/a &  &  & 100\% & 25 \\\hline
\end{tabular}}
    \caption{\textit{Technology data}}
    \label{tab:tech_data}
\end{table}

\begin{table}[H]
    \centering
    \resizebox{0.7\textwidth}{!}{\begin{tabular}{rrrr}
  \hline
  \textbf{Country} & \textbf{Technology} & \textbf{Minimum Capacity (GW)} & \textbf{Maximum Capacity (GW)} \\\hline
   DE & Biomass & 0.0 & 7.57 \\
  ES & Biomass & 0.0 & 1.73 \\
  UK & Biomass & 0.0 & 6.44 \\
  DE & Nuclear & 0.0 & 0.0 \\
  ES & Nuclear & 3.04 & 3.04 \\
  UK & Nuclear & 9.33 & 9.33 \\
  DE & Offshore Wind & 23.63 & 74.25 \\
  ES & Offshore Wind & 0.2 & 17.4 \\
  UK & Offshore Wind & 34.75 & 115.85 \\
  DE & Onshore Wind & 75.37 & 683.67 \\
  ES & Onshore Wind & 48.35 & 1031.77 \\
  UK & Onshore Wind & 26.59 & 542.02 \\
  DE & Run-of-River & 4.73 & 4.73 \\
  ES & Run-of-River & 3.64 & 3.64 \\
  UK & Run-of-River & 2.1 & 2.1 \\
  DE & Solar PV & 96.14 & 8969.8 \\
  ES & Solar PV & 38.4 & 18142.6 \\
  UK & Solar PV & 23.41 & 6612.46 \\\hline
  \end{tabular}
}
    \caption{\textit{Capacity bounds for generation technologies}}
    \label{tab:tech_bounds}
\end{table}

\begin{table}[H]
    \centering
    \resizebox{\textwidth}{!}{\begin{tabular}{rrrrrrrr}
  \hline
    \textbf{Country} & \textbf{Storage Type} & \textbf{Min In (GW)} & \textbf{Max In (GW)} & \textbf{Min Out (GW)} & \textbf{Max Out (GW)} & \textbf{Min Energy (GWh)} & \textbf{Max Energy (GWh)}\\\hline
  DE & Pumped hydro & 7.42 & 7.42 & 7.38 & 7.38 & 242.17 & 242.17 \\
  ES & Pumped hydro & 6.68 & 6.68 & 6.87 & 6.87 & 99.04 & 99.04 \\
  UK & Pumped hydro & 2.68 & 2.68 & 2.95 & 2.95 & 26.38 & 26.38 \\
  DE & Li-Ion & 0.0 & Inf & 0.0 & Inf & 0.0 & Inf \\
  UK & Li-Ion & 0.0 & Inf & 0.0 & Inf & 0.0 & Inf \\
  ES & Li-Ion & 0.0 & Inf & 0.0 & Inf & 0.0 & Inf \\\hline
\end{tabular}
    }
    \caption{\textit{Capacity bounds for storage technologies}}
    \label{tab:sto_bounds}
\end{table}

\begin{table}[H]
    \centering
    \resizebox{0.7\textwidth}{!}{\begin{tabular}{rrrr}
  \hline
  \textbf{Country} & \textbf{Electricity Demand (TWh$_{el}$)} & \textbf{H2 Demand (TWh$_{H2}$)} & \textbf{Mean Heat Demand (TWh$_{th}$)} \\\hline
  DE & 696.3 & 42.08 & 450.27 \\
  ES & 346.14 & 6.68 & 214.72 \\
  UK & 339.81 & 13.81 & 303.99 \\\hline

\end{tabular}
    }
    \caption\textit{{Aggregate energy demands}}
    \label{tab:agg_demand}
\end{table}

\subsection{Stagewise independence test} \label{sec:si_stagewise}

In this section, we describe the approach to computing the autocorrelation functions to justify the stagewise independence and rationalize the foresight assumption. We take four-hourly data for each of the weather variables of interest. For a given weather variable $x$ we compute the monthly average value given by,
\begin{align*}
    \bar{x}_{y,m} := \frac{1}{H_m}\sum_{h}x_{y,m,h}  \quad \forall  m\in\{1,\dots,12\}, y\in\{1982/83,\dots,2016/17\}
\end{align*}

\noindent Next, we de-seasonalize the monthly average inputs by subtracting the average for a given month across all weather years,
\begin{align*}
    \bar{x}^{norm}_{y,m} :=  \bar{x}_{y,m} - \frac{1}{35}\sum_y \bar{x}_{y,m} \quad \forall  m\in\{1,\dots,12\}, y\in\{1982/83,\dots,2016/17\}
\end{align*}

\begin{figure}[H]
    \centering
    \includegraphics[width=\linewidth]{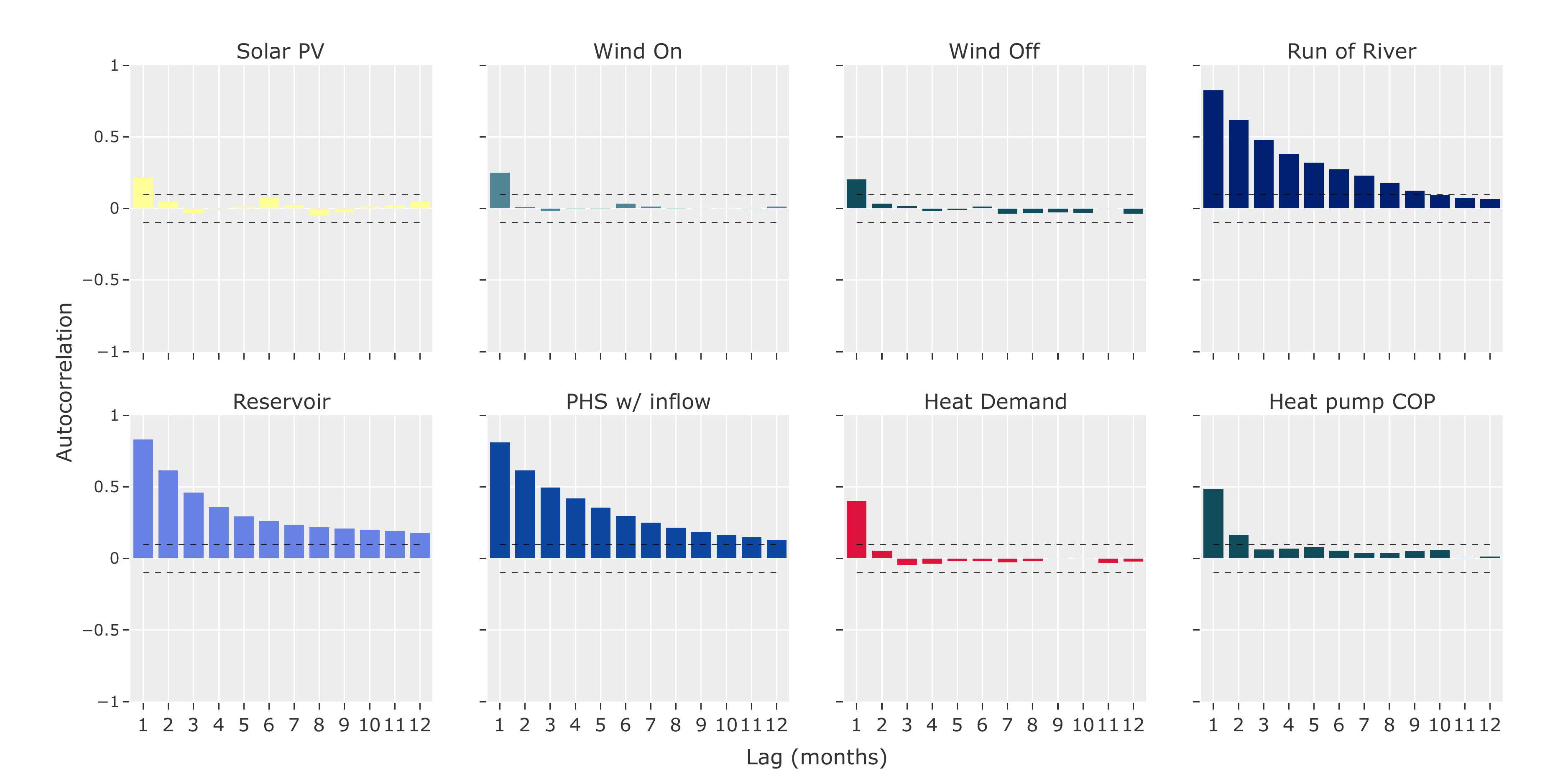}
    \caption{\textit{Autocorrelation functions for weekly stage length}}
    \label{fig:acf_weekly}
\end{figure}

\noindent The resulting time series gives the monthly average input values as deviations from their seasonal mean. Let $T=420$ be the total number of monthly observations. We estimate the autocorrelation function for 12 lags on these values according to,
    \begin{align*}
        \rho_k := \frac{\frac{1}{T-k+1}\sum_m \bar{x}^{norm}_t \bar{x}^{norm}_{t-k}}{\sigma^2}
    \end{align*}
\noindent where $\sigma$ is the standard deviation of the monthly time series. Figure \ref{fig:acf_monthly} and Figure \ref{fig:acf_weekly} plot these autocorrelations as a function of $k$.

\end{document}